\documentclass{emulateapj}

\usepackage{lscape}

\newcommand{\msun}{{\rm M}_{\odot}}
\newcommand{\rsun}{{\rm R}_{\odot}}

\newcommand{\kms}{\rm km\ s^{-1}}

\newcommand{\oneday}{\rm day}
\newcommand{\days}{\rm days}

\newcommand{\myr}{\rm Myr}
\newcommand{\ksubw}{K_{\rm w}}

\newcounter{results_table_num}  

\begin{document}

\shorttitle{Rotational evolution of fully-convective stars}
\shortauthors{Irwin et al.}

\title{On the angular momentum evolution of fully-convective stars:
  rotation periods for field M-dwarfs from the MEarth transit survey}

\author{Jonathan~Irwin, Zachory~K.~Berta, Christopher~J.~Burke,
  David~Charbonneau and Philip~Nutzman}
\affil{Harvard-Smithsonian Center for Astrophysics, 60 Garden St.,
  Cambridge, MA 02138, USA}
\email{jirwin -at- cfa.harvard.edu}

\author{Andrew~A.~West}
\affil{Department of Astronomy, Boston University, 725 Commonwealth Ave, Boston, MA 02215, USA}

\author{Emilio~E.~Falco}
\affil{Fred Lawrence Whipple Observatory, Smithsonian Astrophysical Observatory, 670 Mount Hopkins Road,
  Amado, AZ 85645, USA}

\begin{abstract}
We present rotation period measurements for 41 field M-dwarfs, all of
which have masses inferred (from their parallaxes and 2MASS K-band
magnitudes) to be between the hydrogen burning limit and $0.35\
\msun$, and thus should remain fully-convective throughout their
lifetimes.  We measure a wide range of rotation periods, from $0.28\
\days$ to $154\ \days$, with the latter commensurate with the typical
sensitivity limit of our observations.  Using kinematics as a proxy
for age, we find that the majority of objects likely to be thick disk
or halo members (and hence, on average, older) rotate very slowly,
with a median period of $92\ \days$, compared to $0.7\ \days$ for
those likely to be thin disk members (on average, younger), although
there are still some rapid rotators in the thick disk sample.  When
combined with literature measurements for M-dwarfs, these results
indicate an increase in spin-down times with decreasing stellar mass,
in agreement with previous work, and that the spin-down time becomes
comparable to the age of the thick disk sample below the fully-convective
boundary.  We additionally infer that the spin-down must remove a
substantial amount of angular momentum once it begins in order to
produce the slow rotators we observe in the thick disk candidates,
suggesting that fully-convective M-dwarfs may still experience strong
winds.
\end{abstract}


\section{Introduction}
\label{intro_sect}

The rotational evolution of low-mass stars is predominantly governed
by two competing processes.  During the pre-main-sequence (PMS) phase,
these stars are still collapsing, and thus, the moment of inertia
decreases as a function of time.  If angular momentum is conserved, the
angular velocity must correspondingly increase.  This spin-up, which
persists until the star reaches the zero age main sequence (ZAMS), is
counteracted by angular momentum losses, which are thought to be
related to the star-disc interaction at early times
(e.g. accretion-driven winds; \citealt{mp05}, or ``disc locking'';
\citealt{k91}; \citealt*{cc95}), and stellar winds at late times,
particularly after the star reaches the ZAMS.

It is well-established that for solar-type stars, it is possible to
reproduce the observed evolution reasonably well within this
framework, taking as theoretical inputs a range of disc lifetimes in
reasonable consistency with those observed in young clusters, a wind
loss law involving saturation of the angular momentum losses past some
critical angular velocity $\omega_{\rm sat}$ (e.g., \citealt{sh1987};
\citealt{bs96}), and allowing the radiative core to decouple in
angular velocity from the convective envelope (e.g., \citealt{k97};
\citealt{a98}; \citealt{den2010}).  A key feature of this solar-type
evolution is that the spread in disc lifetimes gives rise to a spread
in rotation rates in young clusters, with the maximal rotation rate
being attained as the stars reach the ZAMS, after which they begin to
spin down, and converge toward a narrow range of rotation rates.
Around the age of the Hyades ($625 \pm 50\ {\rm Myr}$ from isochrone
fitting; \citealt{perryman1998}), the convergence is complete and all
the stars follow a $t^{1/2}$ type spin-down thereafter, with rotation
rate being a well-defined function of mass and age for F, G and K
stars \citep{barnes2003,barnes2007}.

Stellar winds depend on the magnetic dynamo that drives field
generation, so the magnetic topologies of the stars are important in
determining the angular momentum loss rates.  For solar-type stars,
this is thought to be an $\alpha\Omega$ dynamo (where $\alpha$ and
$\Omega$ refer to the ``$\alpha$ effect'', the twisting of the
magnetic field lines caused by rotation, and the ``$\Omega$ effect'',
driven by differential rotation) operating at the interface between
the radiative core and the convective envelope.  For a
fully-convective star, it is thought that this can no longer operate,
and it has been suggested that a turbulent dynamo \citep{durney1993}
or $\alpha^2$ dynamo \citep{radler1990} may dominate.  The former
produces small-scale fields that would yield inefficient angular
momentum losses through winds.

In recent years, magnetic field measurements of stars on both sides of
the fully-convective boundary have become available (e.g.,
\citealt{donati2008,morin2008,morin2010,reiners2009}).  These studies
indicate there is indeed an abrupt change in field geometries moving
across the fully-convective boundary, with fully-convective objects
storing more magnetic flux in large-scale fields than
partially-convective objects.  The effect of these changes on winds is
not yet clear, but at any rate, we do not necessarily expect that the
same wind formalism that works for solar type stars or
partially-convective M-dwarfs should necessarily reproduce the
observations for fully-convective M-dwarfs.

While several authors have attempted to extend the analysis of
rotational evolution to M-dwarfs, and particularly to masses below the
fully-convective boundary ($\approx 0.35\ \msun$;
\citealt{chabrier1997}), the lack of observations in this mass domain
has made such an endeavor problematic.  While observations in young
open clusters have improved dramatically over the last decades, now
yielding large samples of rotation periods below the fully-convective
boundary (e.g.,
\citealt{stassun99,herbst2001,makidon2004,lamm2005,se2004a,se2004b,se2005,se2007,sem2009,cohen2004,little2005,little2010,cieza2006,irwin2007b,irwin2008a}),
these extend typically only to a few 100 $\myr$, and at present only a
handful of periods have been measured for older clusters, where the
intrinsic faintness of M-dwarfs makes determination of periods
difficult.  Furthermore, such objects reach the ZAMS at much later
ages (a few hundred Myr; \citealt{bcah98}) than solar-type stars, so
older clusters are needed to probe similar stages in the evolution.

Additionally, the Sun has been used in all the studies of solar-type
stars as a reference point to tie down the evolutionary models,
particularly the normalization of the wind loss law, which relies on
having data at late-times.  For fully-convective stars, there is as
yet no such object: a star with well known mass, age, and a robustly
measured rotation period (although Proxima comes closest to providing
it; see later in this section).  The best progress below the
fully-convective boundary toward determining the behavior at late
times has been made through $v \sin i$ observations of field stars
(e.g., \citealt{delfosse1998}; \citealt{mohanty2003};
\citealt{reiners2008}; \citealt{browning2010}).  While this
information is invaluable, and has led to important insights into the
physical processes at play in these stars, it is important to note
that $v \sin i$ observations are not sensitive to the slowest
rotators, with typical spectral resolutions yielding a limit of $3\
\kms$.  Due to the small radii of M-dwarfs, this corresponds to rather
short rotation periods, e.g. $3.3\ \days$ for a $0.2\ \rsun$ star.

Rotation period measurements of field M-dwarfs have provided a few
clues as to what might lie below the sensitivity limits of the $v \sin
i$ surveys.  \citet{benedict1998} report a rotation period of $83\
\days$ for Proxima Centauri, a star which is highly likely to be
fully-convective (they also report a weaker detection of a $\approx 130\
\oneday$ periodicity in Barnard's star).  \citet{kiraga2007} presented
a comprehensive survey of rotation periods for field M-dwarfs,
and whilst their sample is limited mostly to objects above the
fully-convective boundary (they present periods for two new objects
below it), they were able to confirm a long period for Proxima,
obtaining $82.5\ {\rm days}$.  \citet{hartman2009b} also report a
number of long rotation periods for field M-dwarfs from the HATNet
transit survey.  Finally, we see evidence for a comparably long
rotation period in the transiting exoplanet host GJ~1214
\citep{charbon2009}.

The MEarth transit survey (\citealt{nutzman2008}; \citealt{irwin2009})
targets nearby ($< 33\ {\rm pc}$) northern hemisphere mid to late
M-dwarfs to search for transiting super-Earth exoplanets in the
habitable zones of their parent stars.  The targets for this survey
are selected to have inferred radii $< 0.33\ \rsun$ because this is
highly advantageous for a transit survey \citep{nutzman2008}, and thus
should all lie below the fully-convective boundary.  Each star is
typically observed at moderate ($20\ {\rm minute}$) cadence on every
clear night for a long period of time ($\approx 6\ {\rm months}$ with the
current implementation of the survey).  The data therefore possess
favorable sampling to search for long rotation periods.

\section{Observations and data reduction}
\label{obs_sect}

MEarth is a targeted survey, and thus, has observational properties
rather different from the majority of transit surveys.  We summarize
the salient features in this section, particularly as they relate to
the detection of continuous photometric modulations, rather than the
discrete transit events for which the survey was designed.

The data for the present work were gathered using all $8$ telescopes
of the MEarth array, which is located within a single roll-off roof
enclosure at the Fred Lawrence Whipple Observatory on Mount Hopkins,
Arizona.  Data were from two full observing seasons, 2008/2009 and
2009/2010, which run from approximately mid-September to mid-July,
where the observatory is shut down during the remaining $\approx 2\ {\rm
  month}$ period corresponding to the summer monsoon season in
southern Arizona.  Each star was generally observed during only one of
these two seasons, as our target list is cycled annually to increase
the sample size searched for transits.  During each year of
operations, hardware changes were minimal except for necessary
repairs, and we used a fixed $715\ {\rm nm}$ long pass filter combined
with a thinned, back illuminated $2048 \times 2048$ CCD on each
telescope.  The pixel scale is $0\farcs76$ per pixel, yielding a
field-of-view of $26'$ on a side.

Targets for the MEarth survey were selected by \citet{nutzman2008}
from a subsample of stars from the LSPM-North catalog \citep{ls2005}
with trigonometric parallaxes or spectroscopic or photometric distance
moduli indicating they are within $33\ {\rm pc}$ \citep{lepine2005}.

For the purposes of this work, it is important to be able to, at a
minimum, assign a reliable estimate of mass or spectral type to each
target, and produce some estimate of the age.  The majority of the
MEarth targets are quite poorly characterized from existing data in
the literature, often possessing only near-IR JHK magnitudes from the
2MASS all-sky survey, photographic magnitudes from the Palomar Sky
Surveys\footnote{Note that although \citet{ls2005} provide $V$-band
magnitude estimates, the majority of these are from photographic plate
measurements from the USNO-B1.0 catalog, and therefore have
potentially large systematic uncertainties.}, and proper motion
information.  We therefore elected to analyze only the $\approx 1/3$
of the sample with trigonometric parallaxes.  There were $273$ such
stars possessing more than one observation (in practice, more
observations are of course needed to detect a period; the smallest
number of measurements for our selected rotation candidates was $207$
for GJ~51, see also \S \ref{sample_sect}).  The combined parallax and
K-band magnitude information provides one of the best methods to
estimate the masses of single field M-dwarfs \citep{delfosse2000}, and
in combination with the proper motion information, can be used on a
statistical basis to constrain the Galactic population to which the
target belongs, and hence the age.  Note that full kinematic
information is not available for the majority of our targets as they
lack radial velocities, so this population assignment is necessarily
quite crude at the present time.

During the night, each telescope observes $20-30$ fields, where
the vast majority of fields contain only a single target M-dwarf.
These are observed sequentially for the entire time they are above
airmass $2$, cycling around the list to yield a cadence per-field of
$\approx 20\ {\rm minutes}$.  Exposure times are selected to yield a
signal to noise ratio for the target star sufficient to detect a $2\
{\rm R_\oplus}$ planet transit at $3 \sigma$ per data point, for the
assumed stellar radius, which is based on the \citet{delfosse2000}
K-band mass-absolute magnitude relation, and a polynomial fit to the
empirical mass-radius data of \citet{ribas2006} to convert mass to
radius.

Basic reductions and light curve generation were performed using
an automated pipeline, based on that used by the Monitor project
\citep{irwin2007a}.  A number of instrument-specific refinements have
been made, and these will be described in full in a forthcoming
publication (Berta et al., in preparation), along with our transit
search procedures.  There are, however, known to be two important
systematic effects remaining after this standard processing, which we
will now describe as they are important for rotation period detection.

MEarth uses German Equatorial Mounts, which necessitate effectively
rotating the telescope through $180^\circ$ relative to the sky when
crossing the meridian.  Thus, each target samples two regions of the
detector, one for negative hour angle and one for positive hour
angle.  Flat fielding errors are manifest as different base-line
magnitudes on each side of the meridian, so we solve for a ``meridian
offset'' for each object to remove this effect.

Secondly, we discovered correlations between the measured differential
magnitudes of the target M-dwarfs and weather parameters, specifically
the relative humidity.  This effect has been investigated in some
detail and will be described by Burke et al. (in preparation).  It
results from a mismatch in spectral type between the target 
star and the comparison stars, where typically the comparison stars
are much bluer than the targets (usually, they are close to solar-type
for the majority of our fields).  Variations in the precipitable water
vapor (PWV) content of the atmosphere along the line of sight cause the
strength of the telluric water vapor absorption in the MEarth bandpass
to vary, which affects the photometry of the targets in a way which is
not corrected by standard differential photometry procedures
\citep{angione1999,bjl2003}.  This effect can reach several percent
over the course of an observing season for the later M-dwarf spectral
types.

Since we lack sufficiently red comparison stars with good signal to
noise ratios for the majority of our target fields, we adopt an
alternative method to derive the required photometric correction.  We
create a ``common mode'' light curve by taking the median of all of
the M-dwarf differential magnitudes observed by all $8$ MEarth
telescopes in a half hour time bin (where the size was chosen to
ensure all stars being observed at a given time are included).  The
averaging is necessary both to improve the signal to noise ratio, and
to remove the effects of any real variability or transits.  Since the
precipitable water vapor usually varies slowly (the majority of the
variations are from night to night) this is sufficient to remove most
of the effect.  The mismatch in spectral type between the target and
the comparison stars varies for each target, and therefore so does the
amplitude of the effect, so in practice we subtract a scaled version
of the ``common mode'', determining the scale factor for each object by
standard least-squares methods.  The scale factor is generally found
to be very well-correlated with the colors (and thus, presumably,
spectral types) of the targets.

\section{Period detection}

\subsection{Method}

In order to properly account for the ``common mode'' systematic effect
and the magnitude offsets between the two sides of the meridian, we
adapt the method described in \citet{irwin2006}, which uses
least-squares fitting of sinusoids to the observed time-series
$m(t)$.

We adopt as the null hypothesis a model of a constant (real)
magnitude, modulated by the systematics corrections:
\begin{equation}
m_0(t) = \left\{ \begin{array}{l}
m_- + k\ c(t), h < 0 \\
m_+ + k\ c(t), h \ge 0 \end{array} \right.
\label{h0_equation}
\end{equation}
where $m_-$ and $m_+$ are separate (constant) baseline flux levels for
the two sides of the meridian, and $k$ is a scale factor multiplying
the ``common mode''.  $h$ is hour angle, and $c(t)$ is the ``common
mode'', determined from a time-binned median of all the M-dwarfs being
observed by MEarth at any given moment.  The scale factor $k$ allows
for the variable amplitude of the ``common mode'' component in each
M-dwarf due to the differing spectral type mismatch of our targets and
comparison stars.

This null hypothesis is compared to the alternate hypothesis that the
light curve contains a sinusoidal modulation, again modulated by the
systematics corrections, of the form:
\begin{equation}
m_1(t) = \left\{ \begin{array}{l}
m_- + k\ c(t) + a \sin(\omega t + \phi), h < 0 \\
m_+ + k\ c(t) + a \sin(\omega t + \phi), h \ge 0 \end{array} \right.
\label{h1_equation}
\end{equation}
where $a$ and $\phi$ represent the semi-amplitude and phase of
the sinusoid, and $\omega = 2 \pi / P$ is the angular frequency
corresponding to rotation period $P$.  In order to determine $\omega$,
we fit this model using standard linear least-squares, at discrete
values of $\omega$ sampled on a uniform grid in frequency from $0.1$
to $1000\ \days$.  By comparing the best-fitting $\chi^2$ values for
the two hypotheses as a function of frequency, we derive a
``least-squares periodogram'' that accounts for the effect of the
systematics.

Due to the small sample size, we omitted the cut in $\chi^2$
improvement used by \citet{irwin2006}, and simply subjected all of the
light curves to visual inspection, to define the final sample of
periodic variables, although in practice all of the objects selected
pass the $\Delta \chi^2_\nu > 0.4$ criterion used by
\citet{irwin2006}.  $41$ light curves passed this selection, where the
remainder were consistent with no detectable variation, had excessive
systematics, or had insufficient data to determine a period.  A number
of objects appeared to have monotonic trends in time, but it is
not clear for many of these if they are due to systematics or to
variability at present.

\subsection{Sample properties}
\label{sample_sect}

\begin{figure}
\centering
\includegraphics[angle=270,width=3.0in]{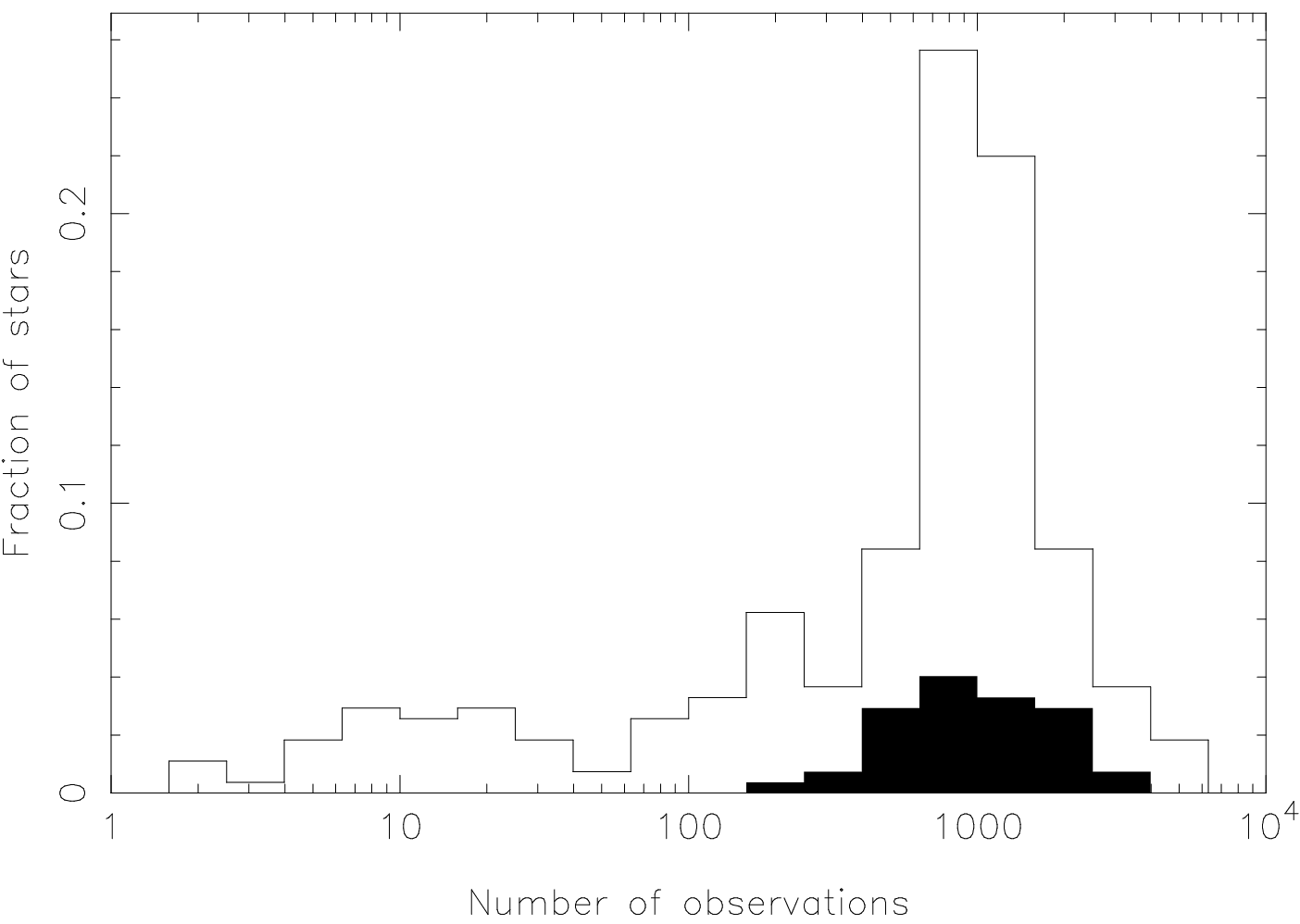}

\vspace{1ex}

\includegraphics[angle=270,width=3.0in]{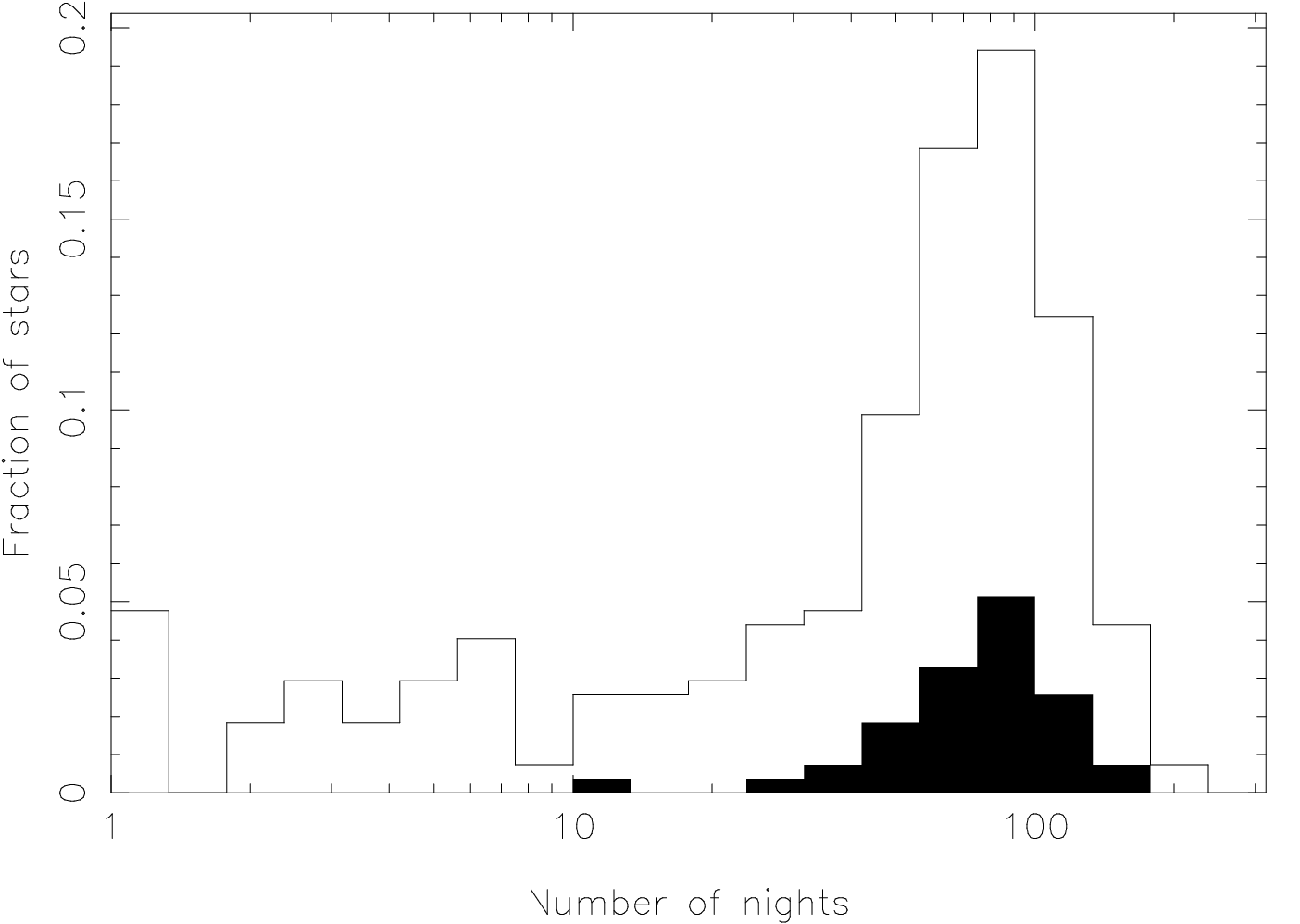}

\vspace{1ex}

\includegraphics[angle=270,width=3.0in]{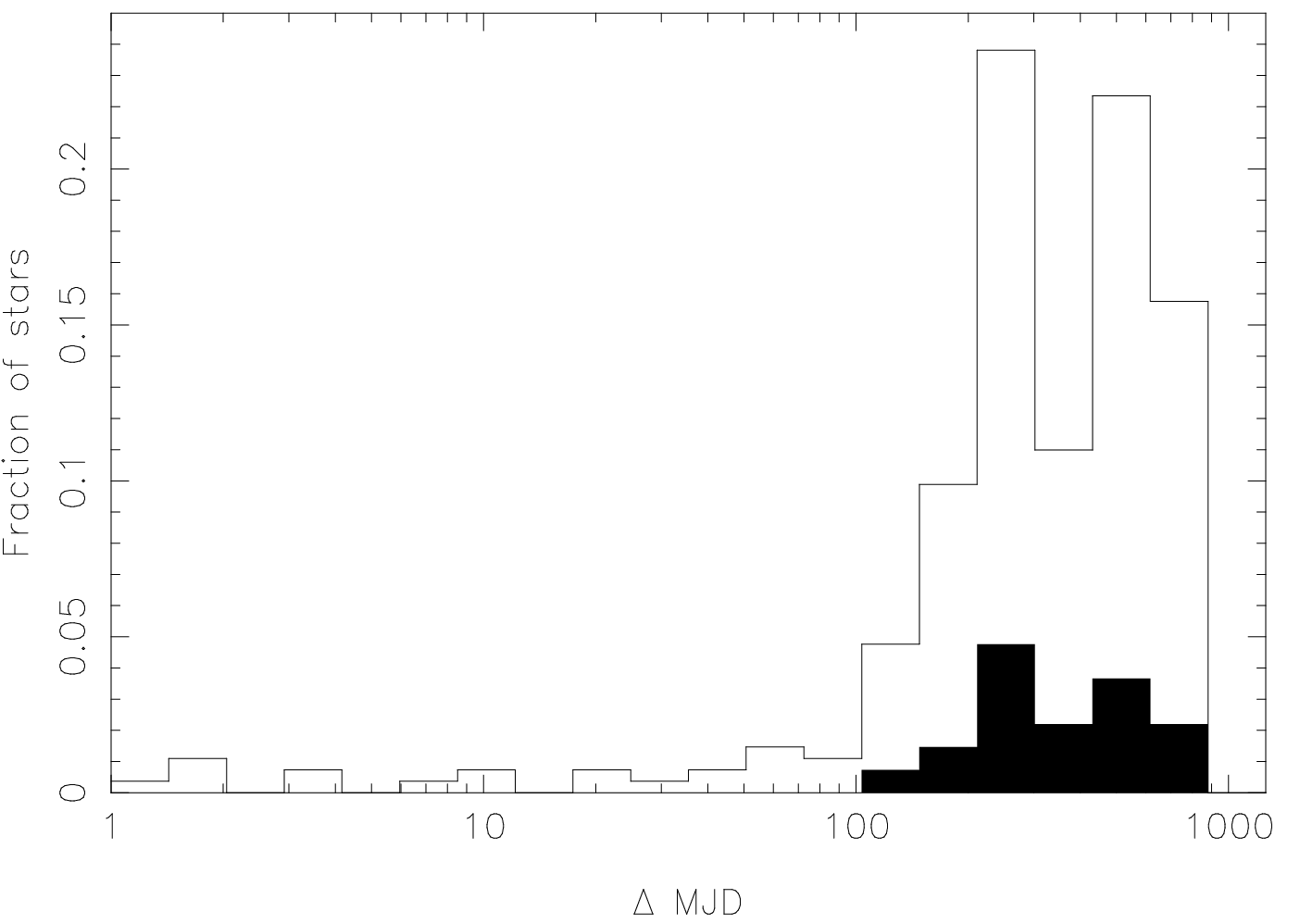}
\caption{Summary of sample properties.  In each panel, the open
  histogram shows the full sample of 273 stars with more than one
  observation, and the solid histogram shows the 41 stars for which we
  detect periods.  The panels show the number of observations for each
  target (top), number of nights during which data were gathered
  (center), and the range of Julian dates spanned by the observations
  (bottom).}
\label{sample}
\end{figure}

Figure \ref{sample} summarizes the overall properties of the sample of
273 stars on which the period search and selection of rotation
candidates was performed.  As discussed in \S \ref{obs_sect}, all
of these stars have trigonometric parallaxes, and Figure \ref{sample}
shows that all of the objects selected as rotation candidates also
have more than $200$ data points taken on $\ge 10$ nights (for all but
one object on $\ge 28$ nights), spanning at least $120$ days.

It is important to note that the selection we have performed does not
remove close binaries from the sample.  Tidal effects will modify the
rotation rates of the components of such binaries, by transferring
angular momentum between the stellar spin and the binary orbit.  The
dependence of tides on binary semimajor axis is very strong
(e.g. \citealt{zahn1977}) so the objects showing the largest tidal
effects will be in the shortest period systems.  It is therefore
possible that some of the objects in the present sample have had their
rotation rates altered by a binary companion, with the objects
rotating at shorter periods being more likely to have been affected.

We also note that the use of K-band absolute magnitude to estimate
masses means that these masses will be overestimated for any
unresolved binary or multiple star systems due to the extra light from
the companion(s).

These issues are best resolved by performing spectroscopic follow-up
at high resolution, to search for double-lined objects and for radial
velocity variability in order to identify any spectroscopic binaries.
This has not yet been done for the present sample, and it is important
to bear the caveats discussed in this section in mind in the
interpretation of the rotation periods we measure.

\subsection{Effect of the ``common mode'' correction}
\label{common_sect}

Since we are solving for the ``common mode'' amplitude simultaneously
with the sinusoidal variability, it is important to evaluate
the effect of this correction on our period sensitivity.

\begin{figure}
\centering
\includegraphics[angle=270,width=3.0in]{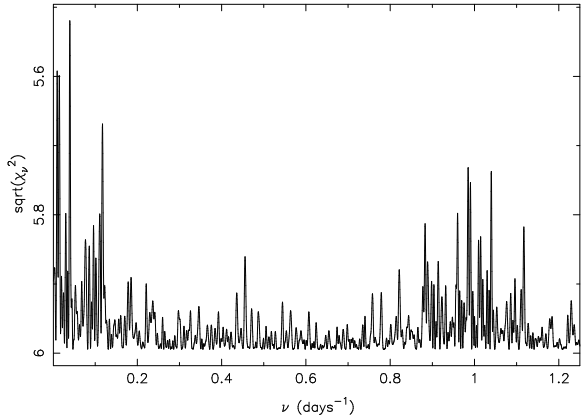}
\caption{Periodogram of the ``common mode'' light curve for the first
  year of observations.  Plotted as the square root of the reduced
  $\chi^2$ of the sinusoidal fit as a function of frequency, with the
  vertical axis inverted such that higher peaks correspond to more
  significant detections as for a conventional periodogram.  The
  period of the highest peak is $25.1\ {\rm days}$.  When computing
  the $\chi^2$ values we used the scatter in each time-bin to compute
  the weights; this appears to significantly underestimate the scatter
  between the bins, hence the large values of the reduced $\chi^2$.
  The peaks around $1\ {\rm day^{-1}}$ appear to be caused by
  aliasing, and probably do not represent real variations in the
  PWV on these timescales.  This has been confirmed by analyzing
  contemporaneous GPS-based PWV measurements (which are not restricted
  to hours of darkness) from Flagstaff, Arizona.}
\label{common-1}
\end{figure}

\begin{figure}
\centering
\includegraphics[angle=270,width=3.0in]{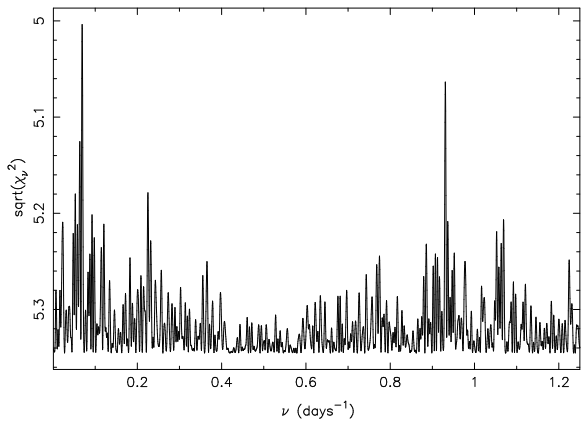}
\caption{As Figure \ref{common-1} only for the second year of
  observations.  The period of the highest peak is
  $14.5\ {\rm days}$.}
\label{common-2}
\end{figure}

Figures \ref{common-1} and \ref{common-2} show periodograms of the
``common mode'' light curve, calculated in the same way as those we
use for period detection by least-squares fitting of sinusoids.  The
dominant power in the ``common mode'' is on long timescales, and at
the corresponding $1\ {\rm day^{-1}}$ alias frequencies, with the
highest peaks for the two years of observations corresponding to
periods of $25.1\ {\rm days}$ for 2008/2009 and $14.5\ {\rm days}$ for
2009/2010.

The presence of power at such frequencies could affect our sensitivity
to rotation periods in this range.  We therefore proceed in the next
section to simulate the full end-to-end period detection process,
including the effect of the ``common mode'' and of the visual
inspection stages, to evaluate the survey sensitivity.

\subsection{Simulations}

In order to evaluate the sensitivity in period and amplitude,
simulations were performed following the method detailed in
\citet{irwin2006}, injecting sinusoids with periods from $0.1$ to $200\
\days$ following a uniform distribution in $\log$ period into only
the light curves rejected in the previous visual inspection stages
to reduce contamination by any real variability.  Two semi-amplitudes
were simulated, $0.005$ and $0.01\ {\rm mag}$, corresponding to the
range of typical amplitudes of our rotation candidates.

\begin{figure}
\centering
\includegraphics[angle=270,width=3.0in]{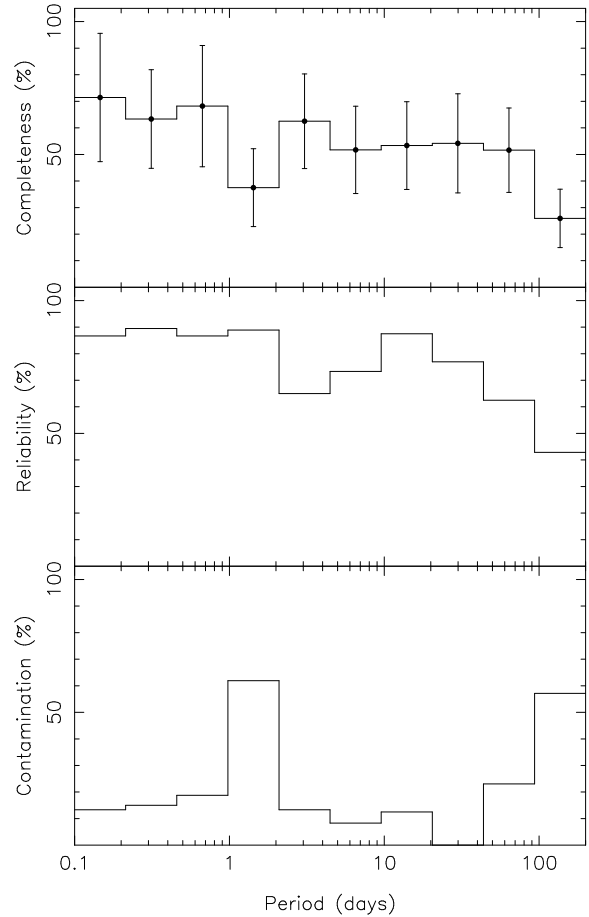}
\caption{Results of the simulations for $0.005\ {\rm mag}$ semi-amplitude
  expressed as percentages, plotted as a function of period.  {\bf Top
  panel}: completeness as a function of real (input) period.  {\bf
  Center panel}: Reliability of period determination, plotted as the
  fraction of objects with a given true period, detected with the
  correct period (defined as differing by $< 10\%$ from the true
  period).  {\bf Bottom panel}: Contamination, plotted as the
  fraction of objects with a given detected period, having a true
  period differing by $> 10\%$ from the detected value.}
\label{sims-0.005}
\end{figure}

\begin{figure}
\centering
\includegraphics[angle=270,width=3.0in]{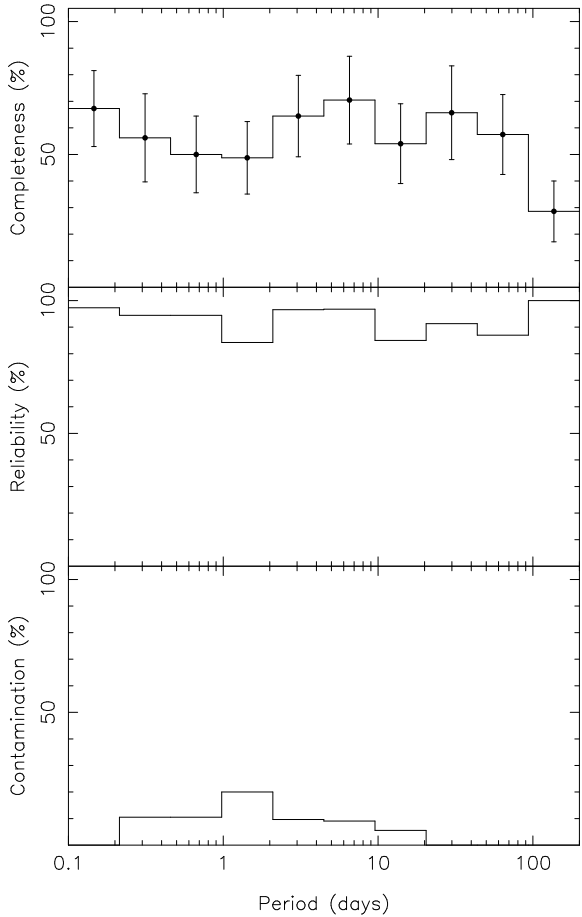}
\caption{Panels as Figure \ref{sims-0.005} except for $0.01\ {\rm
    mag}$ semi-amplitude.}
\label{sims-0.01}
\end{figure}

\begin{figure}
\centering
\includegraphics[angle=270,width=3.0in]{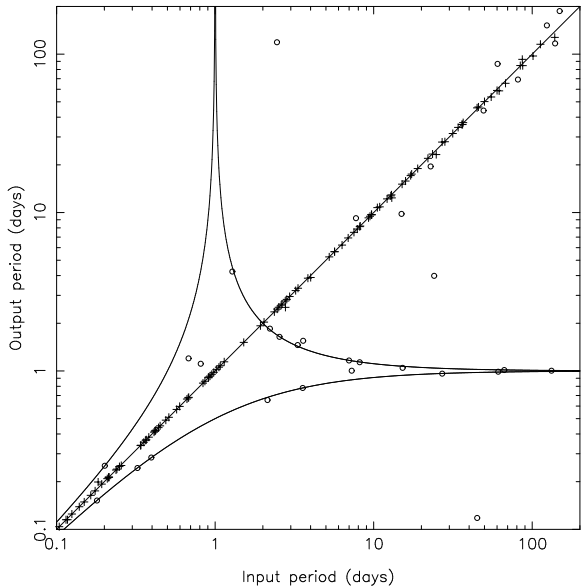}
\caption{Recovered period as a function of actual period for our
  simulations at $0.005\ {\rm mag}$ semi-amplitude.  Crosses represent
  objects where the correct period was recovered to within $10\%$, and
  open circles show objects failing this criterion.  The diagonal
  line indicates equal periods, and the curves show the loci of the
  $1\ {\rm day^{-1}}$ aliases.}
\label{pvsp-0.005}
\end{figure}

\begin{figure}
\centering
\includegraphics[angle=270,width=3.0in]{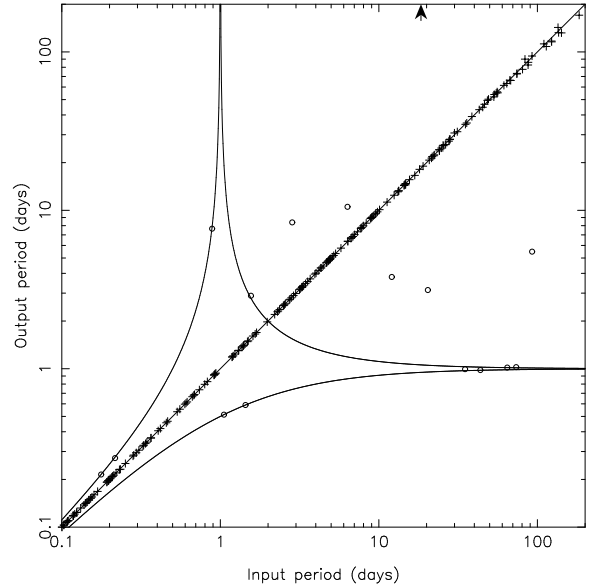}
\caption{Recovered period as a function of actual period for our
  simulations at $0.01\ {\rm mag}$ semi-amplitude.  Symbols and lines
  as Figure \ref{pvsp-0.005}.  The arrow at the top of the diagram
  indicates a single point which fell outside the plotted range.}
\label{pvsp-0.01}
\end{figure}

The results are summarized in Figures \ref{sims-0.005} and
\ref{sims-0.01} for the $0.005$ and $0.01\ {\rm mag}$ semi-amplitudes,
respectively.  We also show plots of the recovered period versus
injected period, in Figures \ref{pvsp-0.005} and \ref{pvsp-0.01},
respectively.  Whilst the overall completeness of rotation period
detections is quite low (around $50-60\%$ in most period bins), there
is no clear bias except in the longest-period bin, where the
completeness drops to $25-30\%$.  This is expected due to the limited
survey duration. 

The reliability and contamination histograms, and Figures
\ref{pvsp-0.005} and \ref{pvsp-0.01} provide an indication of the
reliability of period recovery in the cases where a significant
modulation was detected.  For $0.01\ {\rm mag}$, these statistics
indicate very good period recovery, with reliability (measuring the
fraction of objects detected with the correct period) above $80\%$ in
all period bins, and contamination below $20\%$.  For $0.005\ {\rm
  mag}$ the performance is substantially worse, with a large drop in
reliability around $1\ {\rm day}$ and in the longest-period bin.  The
contamination statistic shows a similar effect.  By examining Figure
\ref{pvsp-0.005}, it is clear that some of the scattering of objects
in and out of the $1\ {\rm day}$ bin is due to aliasing, and this
diagram also indicates that many of the ``incorrect periods''
contributing to the poor performance for the longest periods
simply have larger period errors than our $10\%$ threshold.  This
is probably the reason for the apparently small change in the
completeness statistic between the two bins: this merely counts
detections, without regard to whether the period was correctly
determined.

While it is important to perform the simulation on M-dwarf target
stars to account for the systematics, it is expected that in reality,
many or even the majority of these should show real, astrophysical
variability at some level.  It is therefore likely that we have
injected our simulated signals into objects which also have
astrophysical signals.  For the $0.01\ {\rm mag}$ amplitude, this does
not appear to be a serious problem, with the injected signal generally
overwhelming anything already present.  However, this could contribute
to explaining the apparently high contamination and the lack of a
significant drop in completeness in the $0.005\ {\rm mag}$ amplitude
sample.

\begin{figure}
\centering
\includegraphics[angle=270,width=3.0in]{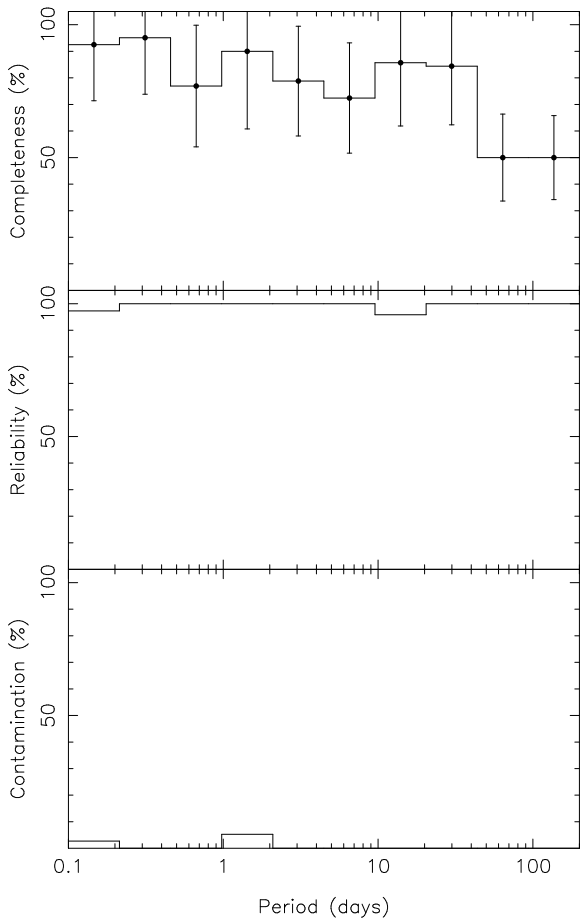}
\caption{Panels as Figure \ref{sims-0.005} except simulating light
  curves with white, Gaussian noise rather than using real data.}
\label{sims-0.005-gauss}
\end{figure}

In order to investigate these effects, and in particular to also test
the influence of systematics on the detection rate, we have performed
an additional set of simulations for $0.005\ {\rm mag}$ semiamplitude
but using white, Gaussian noise of standard deviation set by the
estimated observational errors, rather than using the observed light
curves.  This procedure eliminates systematics (``correlated noise'')
and any existing astrophysical variability but should otherwise have a
comparable noise level to the real data.  The results are shown in
Figure \ref{sims-0.005-gauss}.  This indicates that the low
completeness seen in Figures \ref{sims-0.005} and \ref{sims-0.01} at
short periods results from a roughly equal contribution of systematics
(or contamination from variability) and other effects (sampling and
noise).  The completeness at long periods is slightly improved, but as
expected, is still lower than at short periods, remaining at the
$\approx 50\%$ level.  This is most likely due to the sampling.  The
contamination and reliability histograms show a substantial
improvement, with close to ideal period recovery, which suggests these
effects are almost entirely due to systematics (or contamination from
variability).

\section{Results}

\begin{figure*}
\centering
\includegraphics[angle=0,width=7in]{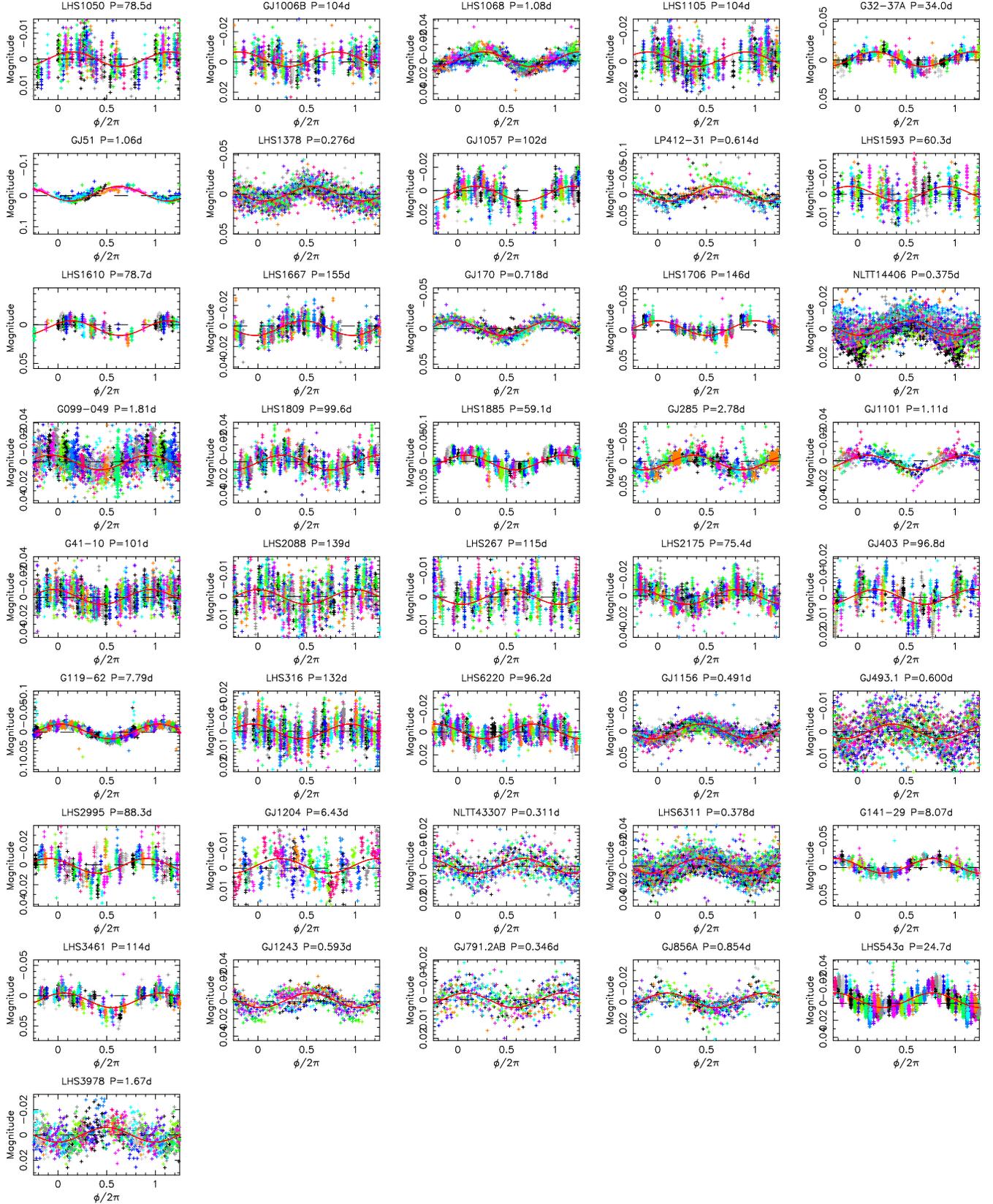}
\caption{Phase-folded light curves for our $41$ rotation candidates.
  Differential magnitudes are plotted, in the MEarth instrumental
  system.  In these diagrams, we plot $\phi/2\pi$ on the horizontal
  axis, where $\phi$ is defined in Eq. (\ref{h1_equation}), and have
  subtracted the ``meridian correction'' and ``common mode
  correction'' described in the text.}
\label{phase_lc}
\end{figure*}

Our results are presented in Table \ref{results_table}, and the
phase-folded light curves of all $41$ objects are shown in Figure
\ref{phase_lc}.



\setcounter{results_table_num}{\thetable}
\addtocounter{table}{1}

It is evident from Figure \ref{phase_lc} that the light curves for
several objects (e.g. LHS 1068, G 32-37A, GJ 170, LHS 543a) are not
perfectly sinusoidal.  This is indeed expected if the modulations
arise from stellar spots.  It is also important to note that there
exist pathological spot distributions (e.g. two identical active
longitudes spaced by $180^\circ$) which can lead to doubling (or in
the general case, multiplication by an integer) of the frequency,
causing us to mis-estimate true the rotation period of the star.  One
way to check for this is to compare the rotation periods with $v \sin
i$ measurements.

We have searched for literature measurements of radial and rotational
velocities for our targets, finding $v \sin i$ measurements for $8$
objects \citep{delfosse1998,mohanty2003,browning2010}.  These are
summarized in Table \ref{vsini_table}, which includes radial
velocities and kinematic population assignments for two additional
objects \citep{montes2001}.

\begin{deluxetable*}{lrrrrrrrrrrrrrr}
\tabletypesize{\normalsize}
\tablecaption{\label{vsini_table} Rotational velocities, radial
  velocities and kinematic populations from the literature for our
  targets}
\tablecolumns{14}

\tablehead{
\colhead{Name} & \colhead{$v \sin i$} & \colhead{$\sigma(v \sin i)$} & \colhead{$v_{\rm rad}$} & \colhead{Population\tablenotemark{a}} & \colhead{Reference\tablenotemark{b}} \\
 & \colhead{($\kms$)} & \colhead{($\kms$)} & \colhead{($\kms$)}
}

\startdata
GJ51       &\ldots  &\ldots &$4.4$  &YD  &1 \\
GJ1057     &$< 2.2$ &\ldots &$27$   &OD  &2 \\
LP412-31   &$12.0$  &$2.0$  &$48.8$ &YD  &3 \\
G099-049   &$7.39$  &$0.8$  &$30$   &YD  &2 \\
LHS1885    &$< 3.7$ &\ldots &$16$   &YO  &2 \\
GJ285      &$4.6$   &$0.4$  &$26.5$ &YD  &4 \\
GJ1156     &$9.2$   &$1.9$  &$4$    &YD  &2 \\
GJ493.1    &$16.8$  &$2.1$  &$-27$  &YD  &2 \\
GJ791.2    &$32.0$  &$2.0$  &$-29$  &YD  &2 \\
GJ856A     &\ldots  &\ldots &$-24$  &YD? &1 \\
\enddata

\tablenotetext{a}{See \citet{leggett1992}.}
\tablenotetext{b}{(1) \citet{montes2001}; (2) \citet{delfosse1998}; (3) \citet{mohanty2003}; (4) \citet{browning2010}}

\end{deluxetable*}

\begin{figure}
\centering
\includegraphics[angle=270,width=3.1in]{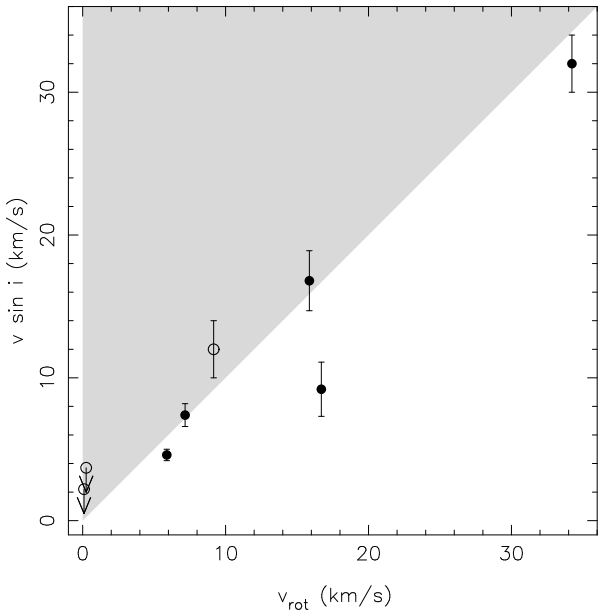}
\caption{$v \sin i$ from spectroscopic measurements plotted as a
  function of rotation velocity from our period measurements, assuming
  the stellar radii given in Table \ref{results_table}.  Points with
  arrows show the two $v \sin i$ limits.  Open symbols denote
  young/old (YO) and old disc (OD) objects, and filled symbols denote
  young disc (YD).  The gray shaded region indicates the part of the
  diagram where $v \sin i > v_{\rm rot}$, and should not be occupied.
  Note that the stellar radii (and thus the estimated rotation
  velocities) also have significant uncertainties, but these are
  difficult to quantify, so we have not plotted horizontal error bars
  on the diagram.  These uncertainties are likely to amount to at least
  $10\%$, and probably greater given the uncertainties in the effects
  of activity on M-dwarf radii (e.g. \citealt{chabrier2007}).}
\label{vsini_comp}
\end{figure}

Figure \ref{vsini_comp} shows a comparison of these $v \sin
i$ measurements with the rotation velocities inferred from our period
measurements and radius estimates.  These are generally in good
agreement, with the exception of one object, LP412-31.  This is one of
the latest-type objects in our sample (\citealt{mohanty2003} state a
spectral type of M8.0), so we suspect the discrepancy may be caused by
an error in the assumed radius when calculating $v_{\rm rot}$,
especially noting the lack of objects at these late spectral types in
the \citet{ribas2006} sample we used to derive the empirical
mass-radius relation.

\subsection{Kinematic population and age assignment}
\label{kin_sect}

It is important for the analysis of rotational evolution (performed
later in this section) to be able to assign approximate ages to the
MEarth sample.  This is a notoriously difficult problem for field
stars.  For the purposes of the present work, we use the available
kinematic information ($5$ of the $6$ phase-space dimensions, where we
lack radial velocities for the majority of the targets) to
statistically assign our targets to the Galactic thin disk or thick
disk / halo populations.

Previous studies have used tangential velocity
(e.g. \citealt{reiners2008}), but this does not take into account the
bulk motions of the various Galactic populations along each
line of sight through the Galaxy.  Instead, we use the model of
\citet{dhital2010} to compute the predicted mean proper motions and
their dispersion for the Galactic thin disk at the position of each
M-dwarf, and compare to the measured proper motions.  Stars within $<
1 \sigma$ of the thin disk prediction were labeled ``thin'' (likely
thin disk members), stars at $> 3 \sigma$ were labeled ``thick''
(corresponding to the thick disk / halo) and are likely to be old, and
the intermediate cases were labeled ``mid''.  These population
assignments are reported in Table \ref{results_table}.

We have assigned a $0.5-3\ {\rm Gyr}$ age to the ``thin'' stars, and
a $7-13\ {\rm Gyr}$ age to the ``thick'' stars (e.g.,
\citealt{feltzing2008}).  The ``mid'' sample in reality contains a
mixture of thin and thick disk populations, so we assign it an
intermediate age of $3-7\ {\rm Gyr}$.

\subsection{Rotation period distribution}

\begin{figure}
\centering
\includegraphics[angle=270,width=3.3in]{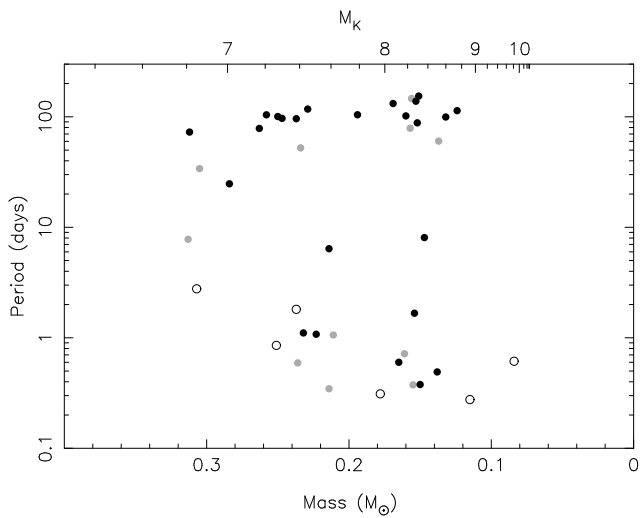}
\caption{Rotation period as a function of stellar mass for our sample.
  The symbols indicate our kinematic population assignments: open
  circles indicate ``thin'' objects (likely thin disk and hence
  younger), gray filled circles ``mid'' objects, and black filled
  circles indicate ``thick'' objects (likely thick disk or halo and
  hence older).}
\label{pmd}
\end{figure}

Figure \ref{pmd} shows rotation period plotted as a function of
stellar mass, where the symbols indicate our kinematic population
assignments from \S \ref{kin_sect}.  This diagram shows no clear trend
of rotation period with stellar mass.  The population appears to be in
two main clumps in the diagram, a population of very rapidly rotating
objects with periods of $\approx 0.2-10\ {\rm days}$, and a population
of very slowly rotating objects with periods of $30\ {\rm days}$ to
the sensitivity limit of the observations at approximately $150\ {\rm
  days}$.

The ``thin'' sample falls entirely within the rapidly rotating clump
of objects, whereas the majority of the ``thick'' stars fall in the
slowly-rotating clump.  The situation is intermediate between these
extremes for the ``mid'' sample.  Given the likely ages of these
populations, this indicates that the older objects in the sample are
on average rotating more slowly.  Such a trend is a natural
consequence of the expectation that these stars spin down due to winds
as they age, and the lack of slowly rotating stars in the ``thin''
sample compared to the abundance of them in the ``thick'' sample
further indicates that this spin-down takes place during ages
intermediate between these two samples for the majority of the
M-dwarfs.  These conclusions confirm the findings of the $v \sin i$
studies (e.g. \citealt{delfosse1998}; \citealt{reiners2008}).

Using a two-sided Kolmogorov-Smirnov test, we have verified that the
difference between the ``thin'' and ``thick'' periods is statistically
significant, with a probability of $4 \times 10^{-3}$ that these are
drawn from the same parent population.  The corresponding
probabilities for the ``thin'' and ``mid'', and the ``mid'' and
``thick'' pairs, are $0.13$ and $0.06$, respectively, neither of which
we consider to be statistically significant.  This is not surprising
as the sample sizes are quite small and the ``mid'' population likely
consists of a mixture of the other two populations.  The median
periods are $0.7$, $7.8$, and $92\ {\rm days}$ for the ``thin'',
``mid'' and ``thick'' samples, respectively.

\section{Discussion}

\subsection{Morphological comparison with open cluster data}
\label{morph_sect}

\begin{figure*}
\centering
\includegraphics[angle=270,width=7in]{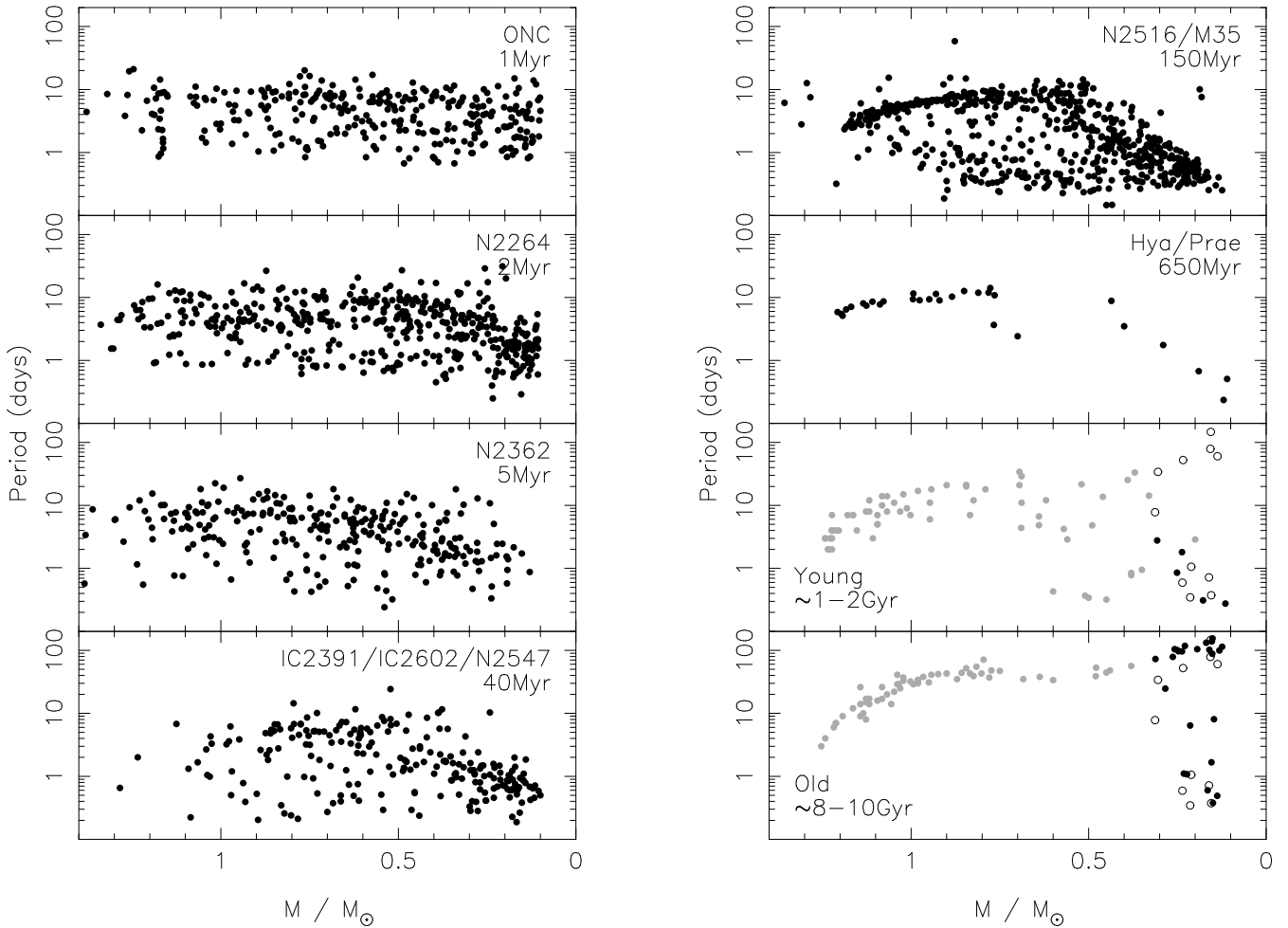}
\caption{Compilation of rotation periods for low-mass stars from the
  literature, including the results from the present study.  Plotted
  in each panel is rotation period as a function of stellar mass for a
  single cluster, or several clusters that are close to coeval.  In
  the case of the field star samples, we divided them into two bins
  for the ``young disc'' and ``old disc'' stars, respectively, and
  color all the literature data gray to distinguish them from the
  results of the present study (in black).  We plot the ``thin''
  population in the ``young'' panel and the ``thick'' population in
  the ``old'' panel, both as solid symbols, with the ``mid''
  population shown in both panels using open symbols (see text). 
  Measurements for field stars are from the compilation by
  \citet{barnes2001,barnes2003} of the rotation periods for the Mount
  Wilson stars from \citet{baliunas1996}, down to $\approx 0.7\ \msun$,
  \citet{kiraga2007} for early-M dwarfs, and the present study for
  later objects.  The appropriate references for each panel are given
  in Table \ref{source_table}.  All the masses for the cluster samples
  were computed using the I-band luminosities of the sources and the
  models of \citet{bcah98}, assuming values of the age, distance
  modulus and reddening for the clusters taken from the literature.
  For the field stars, the \citet{delfosse2000} mass-absolute
  magnitude relations were used for M-dwarfs, and for all higher-mass
  stars, we converted the observed $B-V$ colors to effective
  temperature using Table A5 from \citet{kenyon1995}, and then to mass
  using the models of \citet{bcah98}.}
\label{pmds}
\end{figure*}

In Figure \ref{pmds} (see Table \ref{source_table} for a list of data
sources) we show Figure \ref{pmd} in the wider context of rotation
period distributions from the literature for younger stars in open
clusters, and for more massive stars in the field.

For the MEarth sample, we have assumed the ``thin'' population and
``young disc'' are equivalent, and likewise for ``thick'' and ``old
disc''.  The ``mid'' objects are plotted in both panels using open
symbols as it is not clear to which population these should be
assigned, and it is likely the ``mid'' sample contains members of
both, with the slower rotating objects being more likely to belong to
the ``old disc'' population and vice versa.

\begin{deluxetable}{lll}
\tabletypesize{\normalsize}
\tablecaption{\label{source_table} List of references for rotation
  periods from the literature.}
\tablecolumns{3}

\tablehead{\colhead{Cluster} & \colhead{Age} & \colhead{Source(s)} \\
 & \colhead{(Myr)} &
}

\startdata
ONC      &$1$ &\citet{herbst2001,herbst2002} \\
         &    &\citet{stassun99} \\
\hline
NGC 2264 &$2$ &\citet{lamm2005} \\
         &    &\citet{makidon2004} \\
\hline
NGC 2362 &$5$ &\citet{irwin2008b} \\
\hline
IC 2391  &$30$ &\citet{ps96} \\
IC 2602  &     &\citet{bsps99} \\
NGC 2547 &$40$ &\citet{irwin2008a} \\
\hline
Pleiades &$100$ &\citet{hartman2009a} \\
         &      &\citet{t99} \\
         &      &\citet{se2004b} \\
\hline
NGC 2516 &$150$ &\citet{irwin2007b} \\
M35      &$150$ &\citet{meibom2009} \\
\hline
Hyades   &$625$ &\citet{hya1} \\
         &      &\citet{ple6} \\
Praesepe &$650$ &\citet{se2007} \\
\hline
Field    &      &\citet{baliunas1996} \\
         &      &\citet{barnes2001,barnes2003} \\
         &      &\citet{kiraga2007} \\
         &      &\citet{benedict1998} \\
\enddata

\end{deluxetable}

Figure \ref{pmds} shows that the pre-main-sequence behavior of
M-dwarfs is markedly different from that of solar-type stars.  The
open cluster data indicate that the M-dwarfs all appear to spin up
rapidly, and reach very rapid rotation rates at $100-200\ {\rm Myr}$
age, which corresponds to the time at which the most massive objects
in the $< 0.35\ \msun$ domain reach the ZAMS.  Furthermore, the
slowest rotation period seen at any given mass is itself a very strong
function of mass, declining moving to less massive objects.  The
similar morphologies in the diagrams, especially from NGC 2264 to NGC
2516 (after accounting for the expected spin-up due to contraction,
which corresponds simply to a vertical translation in this logarithmic
plot), indicate there is relatively little rotation rate dependence in
the PMS evolution.

Between the Pleiades age ($\approx 100\ {\rm Myr}$ from isochrone
fitting; \citealt{meynet1993}) clusters and the much older Praesepe
cluster, the available data indicate there is essentially no evolution
of the rotation period, so it appears that little angular momentum is
lost over this phase of the evolution, corresponding to the arrival at
the ZAMS and the early main sequence for these stars.  It is, however,
important to note that the available samples at the Hyades/Praesepe
age are very limited at the present time, so we must caution drawing
strong conclusions from them.  We shall return to this point later in
the discussion.

While the M-dwarfs do not appear to have spun down at the age of the
Hyades/Praesepe, this process is clearly well underway for the earlier
M-dwarfs by the ``young disc'' age, and apparently complete by the
``old disc'' age above approximately $0.4\ \msun$.  Below this mass,
there is a dramatic change moving from the ``young disc'' sample to
the ``old disc'' sample (especially recalling the uncertainty as to
which bin the ``mid'' objects, plotted as open symbols, should be
assigned), where many slowly rotating stars are now found, but there
is still a wide range of rotation periods here, in contrast to the
higher masses, spanning about $2$ orders of magnitude.

It is interesting, and suggestive, to compare the rotation period
distribution in the lower right panel of Figure \ref{pmds} with the
activity fraction (e.g. Fig. 3 of \citealt{west2008}).  The fraction
of active field M-dwarfs shows a steep rise around M5 spectral types.
It is at approximately this point moving down in mass where the rapid
rotation phenomenon appears at the oldest ages.  Given the age range
displayed in the diagram, this seems most likely to be an evolutionary
effect: the time at which the stars spin down increases with
decreasing mass, becoming comparable to the sample age below the
fully-convective boundary.

The existence of a wide range in rotation rates, and the appearance of
a substantial number of slow rotators below the fully-convective
boundary moving between our ``thin'' and ``thick'' bins (although some
are also evident in the ``mid'' bin) further suggests that any spin-down
occurs quite rapidly in the interval between the nominal ages of these
samples for the majority of fully-convective objects, i.e. around $5\
{\rm Gyr}$ in age.

These timescales are in good agreement with those of \citet{west2008},
e.g. their Fig. 10, where they find activity lifetimes of $\approx 7\ 
{\rm Gyr}$ below the fully-convective boundary, and $\la 2\ {\rm Gyr}$
for the partially-radiative objects above it.  We shall explore
possibilities for generating such a spin-down in terms of wind models
in \S \ref{model_sect}.

We note some morphological features in the diagram for the ``old disc''
sample.  The highest mass stars show the familiar break in the
\citet{kraft1967} curve around $1.2\ \msun$, which is thought to be
due to the disappearance of the surface convective zones moving to
higher masses.  Stars without surface convective zones appear to
experience no (or very little) wind losses on the main sequence and
thus remain at very rapid rotation rates until they start to evolve
off the main sequence.

There is another discontinuity in the diagrams around $0.5-0.6\
\msun$, which is highly evident in the NGC~2516 distribution, and has
been noticed and discussed by many authors.  We note here that the
\citet{kiraga2007} sample appears to also show a discontinuity at this
mass for field M-dwarfs, where the rotation period has relatively
little mass dependence immediately above this mass, and appears to
increase with decreasing mass below it.

\begin{figure}
\centering
\includegraphics[angle=270,width=3.2in]{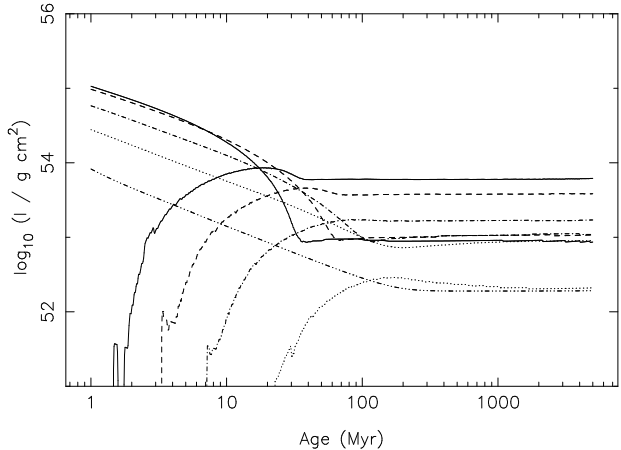}
\caption{Evolution of the moment of inertia from the models of
  \citet{bcah98} for the radiative core (lower lines) and convective
  envelope (upper lines) for masses of $1.0$ (solid line), $0.8$
  (dashed line), $0.6$ (single dot dashed line), $0.4$ (dotted line),
  and $0.2\ \msun$ (triple dot dash line).  Note that the $0.2\ \msun$
  star does not form a radiative core so only one line is shown.}
\label{evol_it_decoup}
\end{figure}

\subsection{Rotational evolution models}
\label{model_sect}

\begin{figure}
\centering
\includegraphics[angle=270,width=3.3in]{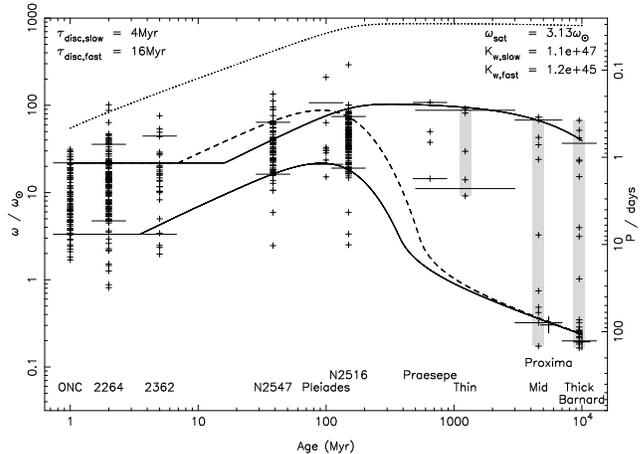}
\caption{Rotational angular velocity $\omega$ plotted as a function of
  time for stars with masses $0.1 < M/\msun \le 0.35$.  Crosses show
  the rotation period data such that each cluster (or field sample)
  collapses into a vertical stripe on the diagram, and short
  horizontal lines show the $10$th and $90$th percentiles of $\omega$,
  used to characterize the slow and fast rotators respectively.
  Plotted are the ONC, NGC 2264, NGC 2362, NGC 2547, the Pleiades, NGC
  2516, Praesepe, and the MEarth data.  We also show the positions of
  Proxima and Barnard's star from \citet{benedict1998} as the large,
  bold crosses.  We have assigned ages to the field stars for the
  purposes of plotting them on this diagram, assuming $6\ {\rm Gyr}$
  for Proxima (the center of the range of ages published for the
  $\alpha$ Cen system; e.g. \citealt{geunther2000};
  \citealt{eggenberger2004}), $10\ {\rm Gyr}$ for Barnard's star, and
  $0.5-3\ {\rm Gyr}$ for the ``thin'', $3-7\ {\rm Gyr}$ for the ``mid'',
  and $7-12\ {\rm Gyr}$ for the ``thick'' MEarth samples (see \S
  \ref{kin_sect}).  These age ranges for the MEarth samples are
  indicated by the lengths of the horizontal lines displaying the
  percentiles, and the MEarth data are shaded gray.  The solid curves
  show rotational evolution models for $0.25\ \msun$ stars, fit to the
  percentiles, Proxima and Barnard's star, with the upper curve for
  the rapid rotators (with parameters $\tau_{\rm disc,fast}$ and
  $K_{\rm w,fast}$) and the lower curve for the slow rotators
  (parameters $\tau_{\rm disc,slow}$ and $K_{\rm w,slow}$).  The
  dashed curve shows the result for the rapid rotators if the wind
  parameter ($K_{\rm w,fast}$) is assumed to be the same as for the
  slow rotators rather than allowing it to vary.  The dotted curve
  shows the break-up limit from Eq. (\ref{pbreak}).}
\label{evol}
\end{figure}

Figure \ref{evol} gives an alternative presentation of the available
rotation period data for objects below $0.35\ \msun$, in a way where
the evolution is more explicit, at the expense of the mass information
seen in the period versus mass plots.  It is important to bear in mind
the strong mass dependence, particularly in the open cluster samples,
when trying to interpret this figure.  We have assigned ages to the
MEarth sample based on the discussion in \S \ref{kin_sect}, noting
that in reality, there is a large spread in age for each of the bins.

In order to model the evolution, we follow the method of
\citet{bouvier1997}.  We adopt a standard wind prescription commonly
used for solar-type stars (e.g., \citealt*{chaboyer1995};
\citealt{bs96}), using a modified version of the \citet{kawaler1988}
wind law with $a = 1$ and $n = 3/2$, introducing saturation of the
angular velocity dependence at a critical angular velocity
$\omega_{\rm sat}$:

\begin{equation}
\left({dJ\over{dt}}\right)_{\rm wind} = \left\{ \begin{array}{r}
-\ksubw\ \omega^3\ \left({R \over{R_\odot}}\right)^{1/2} \left({M \over{M_\odot}}\right)^{-1/2}, \omega < \omega_{\rm sat} \\
-\ksubw\ \omega\ \omega_{\rm sat}^2\ \left({R \over{R_\odot}}\right)^{1/2} \left({M \over{M_\odot}}\right)^{-1/2}, \omega \ge \omega_{\rm sat}
\end{array} \right.
\label{djdt_eqn}
\end{equation}

By coupling this wind prescription to a stellar evolution model, here
the Lyon {\sc NextGen} models of \citet{bcah98}, we model the full
evolution of the angular velocity $\omega$ as a function of time.
This model has several free parameters.  We assume an initial angular
velocity $\omega_0$, here fixed to the observed velocities in the ONC,
and this is held constant for a time $\tau_{\rm disc}$ to account for
disc-related angular momentum losses (``disc locking'') on the
earliest parts of the pre-main-sequence.  The slow and fast rotator
branches are allowed to assume different values of both of these
parameters, with the expectation that the fastest rotators should
decouple from their discs earlier than the slower rotators.  For
the fully-convective objects modeled here, we treat the stars as
solid bodies.  The subsequent evolution is then governed only by the
two parameters $\ksubw$ and $\omega_{\rm sat}$, which control the
normalization of the wind losses and the saturation rate,
respectively.

For a solar-type star, $\ksubw$ can be calibrated by forcing the model to
reproduce the observed rotation rate of the Sun.  For M-dwarfs there
is as yet no analogous object with an old, well-determined age and 
rotation period (although Proxima comes closest to providing it), and
indeed the rotation rates of the slow and fast rotator branches appear
to not yet have converged even by the oldest ages we consider, so
instead we simply fit for this parameter.

Saturation is thought to occur when the rotation period becomes much
smaller than the convective overturn time (e.g., \citealt{k97}).  These 
authors introduced a scaling of $\omega_{\rm sat}$ with mass, of the
form:
\begin{equation}
\omega_{\rm sat} = \omega_{\rm sat,\odot} {\tau_\odot\over{\tau_c}}
\end{equation}
where $\tau_c$ is the convective overturn timescale at $200\ {\rm
  Myr}$.  The convective overturn time lengthens for lower mass stars,
resulting in $\omega_{\rm sat}$ decreasing with decreasing stellar
  mass (corresponding to slower rotation).

Typical values of $\omega_{\rm sat}$ adopted by other authors for the
masses we consider here ($0.25\ \msun$) are of order a few times the
(present-day) solar angular velocity (e.g., \citealt*{sills2000}),
corresponding to periods of a few to ten days.

It is important to note that the convective overturn timescales are
not well-known for fully-convective stars; \citet{sills2000}
extrapolated the theoretical values from \citet{kim1996} to these
masses, and found that the resulting values of $\omega_{\rm sat}$
caused too much angular momentum to be lost compared to the
observations in the Hyades, which were better reproduced with a
smaller value of $\omega_{\rm sat}$.  More recent studies of empirical
convective overturn timescales \citep{pizzolato2003,kiraga2007}
indicate that $\tau_c$ may increase sharply around the
fully-convective boundary, meaning such an extrapolation might
under-predict $\tau_c$, and thus overpredict $\omega_{\rm sat}$, as
found.  It is also possible that the Rossby number is no longer the
dominant factor in determining the operation of the magnetic dynamo
driving stellar winds in fully-convective objects
(e.g. \citealt{mohanty2003,reiners2007}).

The uncertainties associated with the Rossby scaling have led us to
consider several possibilities for $\omega_{\rm sat}$.  We first
consider a value close to that of \citet{sills2000}.  We then allow
essentially free adjustment of $\omega_{\rm sat}$ (\S
\ref{low_omegasat_sect}), and finally consider a third possibility of
there being no un-saturated regime, for comparison, in \S
\ref{zero_omegasat_sect}.

\subsubsection{High $\omega_{\rm sat}$ model}
\label{high_omegasat_sect}

If we start by enforcing values of $\omega_{\rm sat}$ of a few
$\omega_\odot$, the model shown in Figure \ref{evol} results.  We have
performed two fits, one enforcing the same value of $\ksubw$ for both the
fast and slow rotator branches, as would be expected in the saturation
formalism for solar-type stars, and one allowing different values of
$\ksubw$.  We consider first the former model, enforcing the same value of
$\ksubw$ for both branches.

This model is able to reproduce the cluster slow rotators fairly well,
and the oldest ages (Proxima, Barnard's star, and the ``thick'' MEarth
sample).  The disagreement with the Praesepe sample and the ``thin''
sample is more concerning, although we note that the age range of the
latter in practice means it could contain objects as young as the
Hyades/Praesepe, and that the ``thick'' and ``mid'' samples could
contain objects as young as a few Gyr as well.  This model does not
reproduce the rapid spin-down at late times discussed in \S
\ref{morph_sect}.

In the case of the Hyades/Praesepe this model would predict that there
should be objects rotating as slowly as the sun by this age.  There
are to our knowledge no such objects reported in the literature at the
present time, but we suspect the majority of cluster studies carried
out at these masses and ages are not sensitive to periods this long.
For example, \citet{se2007} state that their sensitivity to periods
longer than 1 week is limited.

The situation for the rapid rotators is much less satisfactory, with
difficulties reproducing the data during both the early-PMS and the
main sequence phases.  We first examine the latter.

As for solar-type stars, the wind prescription we have used produces a
rapid convergence in rotation rates at late times.  This does not seem
to be the case for M-dwarfs.  If we instead allow a different value of
$\ksubw$ for the fast rotators, corresponding to allowing the normalization
of the wind law to be rotation rate dependent, it is possible to much
better reproduce the observations from the age of NGC 2516 onwards.
The value of $\ksubw$ found here is two orders of magnitude smaller than
that for the slow rotator branch: these objects experience much weaker
winds in our model.

The theoretical interpretation of such a change likely lies in the
dynamo mechanism operating in these objects, as mentioned in \S
\ref{intro_sect}.  Of the possibilities mentioned there, the
$\alpha^2$ dynamo would possess a strong rotation rate dependence, for
example.

We now return to the question of the early-PMS evolution of the rapid
rotators.  Figure \ref{evol} shows that, contrary to the slow
rotators, these objects appear to spin up more slowly than would
predicted simply from contraction.  Since the stellar evolution model
we are using has no rotation rate dependence, our model would only be
able to reproduce such an evolution by losing angular momentum at
early-times, in this case by choosing a very long disc coupling time
of $16\ {\rm Myr}$.  This seems physically unreasonable (and still
does not fit the data), especially recalling that the rapid rotators
are thought to be produced from stars which uncoupled from their discs
early.

It is possible the prescription for disc locking we have used, of a
constant angular velocity, is not correct, and a more gradual angular
momentum removal operates.  This could result in reduced (rather than
eliminated) spin-up.  The slow rotators do not seem to support this,
showing apparently relatively little evolution as measured by the
percentiles out to ages of $5\ {\rm Myr}$.  It is however important to
note that the slow rotation end of the samples is subject to
contamination from field objects, so it is possible this conclusion is
a result of the higher level of field contamination in the NGC 2362
sample, compared to the ONC and NGC 2264.

Another possible clue may be the proximity of these objects to the
break-up limit.  \citet{herbst2001} give the following formula for the
period corresponding to break-up:
\begin{equation}
P_{\rm break} = 0.116\ {\rm days}\ {(R / {\rm R_{\odot}})^{3/2}\over{(M / {\rm M_{\odot}})^{1/2}}}
\label{pbreak}
\end{equation}
where $M$ and $R$ are the stellar mass and radius, respectively.  This
is plotted as the dotted line on Figure \ref{evol}.  As stated by
\citet{herbst2001}, many of these objects are indeed rotating very
close to the break-up rate, with many approaching $30\%$ of break-up.
This appears to persist to ages of approximately $10\ {\rm Myr}$.

In reality, the proximity of these objects to the break-up limit
means the non-rotating stellar evolution model we have used may no
longer be applicable, because it implies centrifugal forces are likely
to play an important role in determining the stellar structure.  It
may therefore be necessary to invoke a more complete treatment of
stellar evolution, incorporating rotation, to reproduce the behavior of
these objects, for example the treatment used by \citet{sills2000} and
\citet{den2010}.

\subsubsection{Low $\omega_{\rm sat}$ model}
\label{low_omegasat_sect}

\begin{figure}
\centering
\includegraphics[angle=270,width=3.3in]{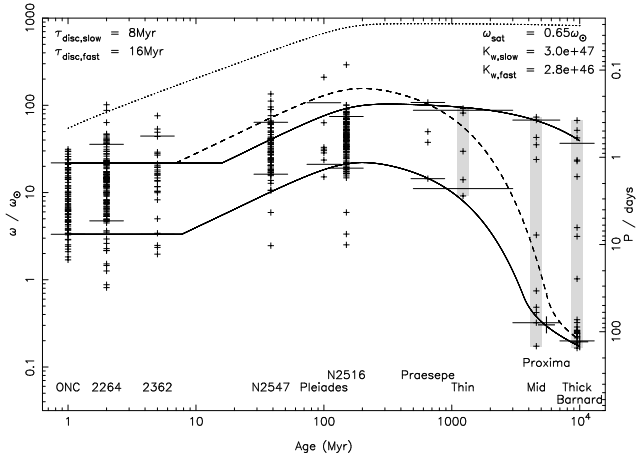}
\caption{As Figure \ref{evol}, except for low values of $\omega_{\rm
    sat}$.}
\label{evol_low_omegasat}
\end{figure}

If we relax the assumption made in \S \ref{high_omegasat_sect} of a
value of $\omega_{\rm sat}$ of a few $\omega_\odot$, it is possible to
obtain an alternate solution which better-fits the behavior at late
times, at the cost of worse agreement with the open cluster data
around the age of the Pleiades.  We show this solution in Figure
\ref{evol_low_omegasat}, which has $\omega_{\rm sat} = 0.65\
\omega_\odot$.

This model appears more satisfactory overall, especially considering
the age ranges in the field samples, and the strong mass-dependence
and possible field contamination in the open cluster samples, meaning
some scatter for the slow rotators in such a wide mass bin is not
entirely unexpected.  The values of the wind parameter $\ksubw$ still
differ for the fast and slow rotator branches, but now by only one
order of magnitude.  The rapid spin-down at late times discussed in \S
\ref{morph_sect} is now produced.

\subsubsection{Zero $\omega_{\rm sat}$ model}
\label{zero_omegasat_sect}

\citet{reiners2008} have offered an interpretation of the available $v
\sin i$ data in terms of a saturated wind braking law, with a
temperature dependence yielding weaker angular momentum losses for
lower-mass objects.  In their analysis, they assume all the objects
rotate faster than the saturation rate, corresponding to elimination
of the $dJ/dt \propto -\omega^3$ branch in Eq. (\ref{djdt_eqn}).  This
could also correspond, for example, to a different magnetic dynamo
than the solar-type objects have at low rotation rates.

After the object arrives on the main sequence, the moment of inertia
and temperature are approximately constant for a given star, so this
angular momentum loss prescription can be trivially integrated to show
that the angular velocity is an exponential function of time.  This
drives the object very quickly to extremely low rotation rate.  In the
context of Figure \ref{evol}, where both axes are logarithmic, this
law corresponds to an exponential decrease in $\log \omega$ versus
$\log t$.

\begin{figure}
\centering
\includegraphics[angle=270,width=3.3in]{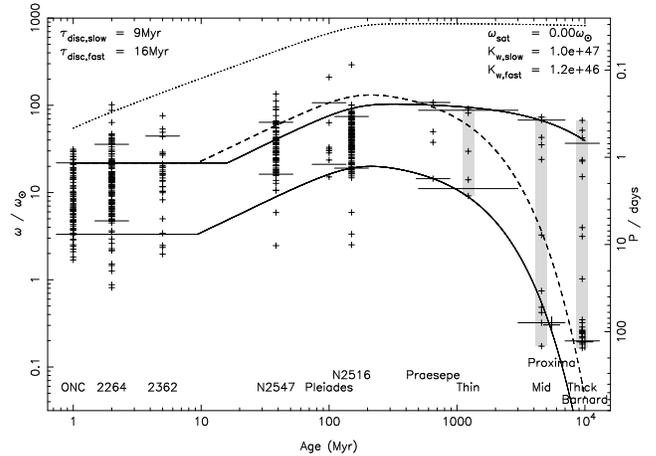}
\caption{As Figure \ref{evol}, except assuming all objects remain in
  the saturated regime throughout their evolution.}
\label{evol_no_omegasat}
\end{figure}

Nonetheless, given the limited quantity of data available it is
possible to obtain a somewhat satisfactory fit using this model, as
shown in Figure \ref{evol_no_omegasat}.

\subsubsection{Overall comments}

The first two models we have considered in this section both exhibit a
non-saturated spin-down at late-times on the slow rotator branch, with
the periods evolving following a $t^{1/2}$ law as for solar-type
stars.  While we have essentially enforced the solar-type law by our
choice of wind prescription, it is important to note that, in reality,
different power laws may be possible depending on the exact magnetic
geometry.  Nonetheless this argues in favor of a phase of
``non-saturated'' angular momentum losses.  Some observational
evidence in favor of fully-convective M-dwarfs being able to support a
non-saturated dynamo exists in the form of a possible
rotation-activity correlation (e.g., \citealt{delfosse1998};
\citealt{mohanty2003}; although see also \citealt{west2009}).

We note that the period of $\approx 130\ {\rm days}$ measured for
Barnard's star by \citet{benedict1998} may also argue in favor of a
$t^{1/2}$ or similarly slow spin-down at late times, and thus against
the model in \S \ref{zero_omegasat_sect}.  Although the evidence for
such a periodicity is stated by \citet{benedict1998} themselves as
being weak, the MEarth sample contains similarly long periods.
Barnard's star appears to be a thick disc object, and thus is likely
to have an age of approximately $7-13\ {\rm Gyr}$ (e.g.,
\citealt{feltzing2008}).  If we assume Proxima and Barnard's star can
be treated as evolutionary analogs, the ratio of rotation periods of
$0.64$ is remarkably close to the predicted $1/\sqrt{2} \simeq 0.71$
from a $t^{1/2}$ law if Barnard's star was twice as old as Proxima.
The exponential spin-down in Figure \ref{evol_no_omegasat} would
produce a much larger change in period.

Finally, the ``pile-up'' of objects at long periods seen in the MEarth
data is difficult to reproduce with an exponential time-dependence in
the late-time evolution.  The convergence in rotation rates produced
by the transition to a $t^{1/2}$ type spin-down law more naturally
produces this feature, whereas the exponential spin-down instead
produces a divergence in periods at late-times when we use the wind
parameter variation with rotation rate in \S
\ref{zero_omegasat_sect}.

Given these lines of reasoning, we believe the model in \S
\ref{zero_omegasat_sect} is unlikely, and further has the somewhat
unsatisfactory feature of producing extreme slow rotators ($\ga 1000\
{\rm days}$) at the oldest ages (see also the discussion in
\citealt{reiners2008}).

Of the remaining two models, we suggest that the observed spin-down
at late-times (see \S \ref{morph_sect}) and the activity lifetimes 
\citep{west2008} support the model presented in \S
\ref{low_omegasat_sect}.  A comparison of Figures \ref{evol} and
\ref{evol_low_omegasat} indicates that the largest differences between
these two possibilities (noting that the age range and small number statistics for
the ``thin'' sample does not strongly constrain the model at present)
are manifest at the age of the Hyades and/or Praesepe, with one model
predicting the existence of objects rotating as slowly as the sun,
and the other being consistent with the presently-available
observations.  More generally, the slowest rotation period seen at
this age constrains the value of $\omega_{\rm sat}$.  We believe these
arguments motivate extending the observations to search for slowly
rotating M-dwarfs in these clusters.

\section{Conclusions}

We have presented a sample of rotation period measurements for 41
field M-dwarfs inferred to lie below the fully-convective boundary,
from the MEarth transit survey.  We measure a wide range of periods,
from $0.28\ {\rm days}$ to $154\ \days$, where the latter is
commensurate with the sensitivity limit resulting from the typical
baseline of the observations, finding that the kinematically young
(thin disk) objects rotate faster than the kinematically old (thick
disk) objects.

Combining the available measurements from the literature with the
present sample shows two interesting features in the mass dependence
of rotation for M-dwarfs at the oldest field ages.  One is the
familiar discontinuity at approximately $0.6\ \msun$ seen in many open
cluster studies, where the mass-dependence of the rotation periods
steepens below this mass.  Contrary to the findings of the open cluster
studies, the field stars below $0.6\ \msun$ appear to rotate {\it
  slower} with decreasing mass.

The second feature is the onset (moving down in mass) of rapid
rotation at old ages for some objects below $0.35\ \msun$, which is
also seen in the $v \sin i$ surveys.  Our rotation periods confirm
that not all such field M-dwarfs are rapid rotators, and there exists
a significant, or even dominant, population of slow rotators that
appears between the ages corresponding to our thin and thick disk
samples.  In addition to this confirmation of the results of $v \sin
i$ studies (e.g. \citealt{delfosse1998}), we can now estimate the
rotation rates for the slow rotators, finding them to rotate
approximately two orders of magnitude slower than the remaining
rapidly rotating objects at the age corresponding to our thick disk
sample.  There is no discontinuity in the rotation rates crossing
$0.35\ \msun$ for the slow rotator branch at the oldest ages.

Comparing the results for the partially-radiative and fully-convective
objects indicates that the spin-down time increases moving to lower
masses.  The ``young disk'' sample from \citet{kiraga2007} has a
comparable fraction of rapid rotators to the ``mid'' and ``thick''
samples from the present survey, indicating that the spin-down times
for these populations are comparable to the respective sample ages:
the ``young disk'' sample age for the partly-radiative stars (a few
Gyr) and the thick / ``old disk'' (probably $> 5\ {\rm Gyr}$) for the
fully-convective stars.

We have explored wind models and were able to reproduce these features
by adopting a very low saturation velocity within a standard
\citep{chaboyer1995,bs96} wind prescription, although it was necessary
to assume a rather large rotation rate dependence of the wind
parameter.  In reality, this additional rotation rate dependence
probably corresponds to modifying the power law exponent $n$ in the
wind model, and thus the magnetic field geometry, although it is
beyond the scope of this paper to speculate on such a change.

\acknowledgments The MEarth team gratefully acknowledges funding from
the David and Lucile Packard Fellowship for Science and Engineering
(awarded to DC).  This material is based upon work supported by the
National Science Foundation under grant number AST-0807690.  We thank
Saurav Dhital for computing the Galactic model results along the
sight-lines to our M-dwarfs, Isabelle Baraffe for providing the
stellar evolution model tracks used in \S \ref{model_sect} and for
discussion of the location of the fully-convective boundary as a
function of time, and Jerome Bouvier, Marc Pinsonneault and Alexander
Scholz for assistance with the rotational evolution models and for
helpful discussions regarding the interpretation.  The anonymous
referee is also thanked for a helpful report that significantly
improved the manuscript.  The MEarth team is greatly indebted to the
staff at the Fred Lawrence Whipple Observatory for their efforts in
construction and maintenance of the facility, and would like to
explicitly thank Wayne Peters, Ted Groner, Karen Erdman-Myres, Grace
Alegria, Rodger Harris, Bob Hutchins, Dave Martina, Dennis Jankovsky
and Tom Welsh for their support.

This research has made extensive use of data products from the Two
Micron All Sky Survey, which is a joint project of the University of
Massachusetts and the Infrared Processing and Analysis Center /
California Institute of Technology, funded by NASA and the NSF, NASA's
Astrophysics Data System (ADS), and the SIMBAD database, operated at
CDS, Strasbourg, France.


\setcounter{table}{\value{results_table_num}}

\clearpage
\LongTables
\begin{landscape}
\begin{deluxetable*}{lrrrrrrrrrrrrrr}
\tabletypesize{\normalsize}
\tablecaption{\label{results_table} Rotation periods, kinematics and stellar parameters for the MEarth sample.}
\tablecolumns{15}

\tablehead{
\colhead{Name} & \colhead{$\alpha$\tablenotemark{a}} & \colhead{$\delta$\tablenotemark{a}} & \colhead{Epoch\tablenotemark{a}} & \colhead{$\mu_\alpha$\tablenotemark{b}} & \colhead{$\mu_\delta$\tablenotemark{b}} & \colhead{$\pi$} & \colhead{$\sigma(\pi)$} & \colhead{Ref.\tablenotemark{c}} & \colhead{$K$\tablenotemark{d}} & \colhead{$M$\tablenotemark{e}} & \colhead{$R$\tablenotemark{f}} & \colhead{Period\tablenotemark{g}} & \colhead{$a$\tablenotemark{h}} & \colhead{Pop\tablenotemark{i}} \\
 & \colhead{ICRS} & \colhead{ICRS} & \colhead{(yr)} & \colhead{(\arcsec/yr)} & \colhead{(\arcsec/yr)} & \colhead{(\arcsec)} & \colhead{(\arcsec)} & & \colhead{(mag)} & \colhead{($\msun$)} & \colhead{($\rsun$)} & \colhead{(days)} & \colhead{(mag)} &
}

\startdata
LHS1050    &$00^h15^m49^s.2$ &$+13^\circ33\arcmin22\arcsec$ &$1998.7$ &$ 0.621$ &$ 0.333$ &$0.0866$ &$0.0134$ &1 &$ 7.83$ &$0.26$ &$0.28$ &$ 78.5$ &$0.0028$ &Thick \\
GJ1006B    &$00^h16^m16^s.1$ &$+19^\circ51\arcmin52\arcsec$ &$1998.7$ &$ 0.709$ &$-0.748$ &$0.0660$ &$0.0016$ &2 &$ 8.12$ &$0.26$ &$0.28$ &$  104$ &$0.0049$ &Thick \\
LHS1068    &$00^h24^m34^s.8$ &$+30^\circ02\arcmin30\arcsec$ &$1997.9$ &$ 0.580$ &$ 0.028$ &$0.0528$ &$0.0044$ &1 &$ 8.91$ &$0.22$ &$0.24$ &$ 1.08$ &$0.0090$ &Thick \\
LHS1105    &$00^h35^m53^s.6$ &$+52^\circ41\arcmin37\arcsec$ &$1998.9$ &$ 0.787$ &$-0.186$ &$0.0621$ &$0.0091$ &1 &$ 9.81$ &$0.19$ &$0.21$ &$  104$ &$0.0048$ &Thick \\
G32-37A    &$00^h39^m33^s.5$ &$+14^\circ54\arcmin19\arcsec$ &$1997.8$ &$ 0.321$ &$ 0.039$ &$0.0353$ &$0.0018$ &1 &$ 9.14$ &$0.30$ &$0.32$ &$ 34.0$ &$0.0095$ &Mid \\
GJ51       &$01^h03^m19^s.7$ &$+62^\circ21\arcmin56\arcsec$ &$1999.0$ &$ 0.739$ &$ 0.086$ &$0.0955$ &$0.0073$ &1 &$ 7.74$ &$0.21$ &$0.23$ &$ 1.06$ &$0.0236$ &Mid \\
LHS1378    &$02^h17^m09^s.9$ &$+35^\circ26\arcmin33\arcsec$ &$1998.0$ &$ 0.549$ &$-0.266$ &$0.0964$ &$0.0011$ &1 &$ 9.03$ &$0.12$ &$0.14$ &$0.276$ &$0.0093$ &Thin \\
GJ1057     &$03^h13^m23^s.0$ &$+04^\circ46\arcmin29\arcsec$ &$2000.7$ &$ 1.749$ &$ 0.084$ &$0.1171$ &$0.0035$ &3 &$ 7.85$ &$0.16$ &$0.18$ &$  102$ &$0.0062$ &Thick \\
LP412-31   &$03^h20^m59^s.7$ &$+18^\circ54\arcmin23\arcsec$ &$1997.8$ &$ 0.351$ &$-0.259$ &$0.0689$ &$0.0006$ &4 &$10.66$ &$0.08$ &$0.11$ &$0.613$ &$0.0188$ &Thin \\
LHS1593    &$03^h47^m20^s.9$ &$+08^\circ41\arcmin46\arcsec$ &$2000.9$ &$ 0.465$ &$-0.664$ &$0.0795$ &$0.0035$ &1 &$ 9.04$ &$0.14$ &$0.16$ &$ 60.3$ &$0.0033$ &Mid \\
LHS1610    &$03^h52^m41^s.7$ &$+17^\circ01\arcmin06\arcsec$ &$1997.8$ &$ 0.433$ &$-0.644$ &$0.0700$ &$0.0138$ &1 &$ 8.07$ &$0.16$ &$0.18$ &$ 78.8$ &$0.0099$ &Mid \\
LHS1667    &$04^h22^m33^s.5$ &$+39^\circ00\arcmin44\arcsec$ &$1999.9$ &$ 0.584$ &$-0.603$ &$0.0534$ &$0.0047$ &1 &$ 9.69$ &$0.15$ &$0.17$ &$  154$ &$0.0071$ &Thick \\
GJ170      &$04^h30^m25^s.3$ &$+39^\circ51\arcmin00\arcsec$ &$1998.9$ &$ 0.271$ &$-0.575$ &$0.0959$ &$0.0028$ &1 &$ 8.26$ &$0.16$ &$0.18$ &$0.718$ &$0.0103$ &Mid \\
LHS1706    &$04^h50^m50^s.8$ &$+22^\circ07\arcmin22\arcsec$ &$1997.8$ &$ 0.636$ &$-0.433$ &$0.0711$ &$0.0057$ &1 &$ 8.99$ &$0.16$ &$0.18$ &$  147$ &$0.0121$ &Mid \\
NLTT14406  &$05^h03^m05^s.6$ &$+21^\circ22\arcmin36\arcsec$ &$1997.8$ &$ 0.109$ &$-0.140$ &$0.0365$ &$0.0084$ &1 &$ 8.91$ &$0.15$ &$0.18$ &$0.375$ &$0.0050$ &Mid \\
G099-049   &$06^h00^m03^s.5$ &$+02^\circ42\arcmin24\arcsec$ &$1999.8$ &$ 0.311$ &$-0.042$ &$0.1862$ &$0.0101$ &3 &$ 6.06$ &$0.24$ &$0.26$ &$ 1.81$ &$0.0076$ &Thin \\
LHS1809    &$06^h02^m29^s.2$ &$+49^\circ51\arcmin56\arcsec$ &$2000.1$ &$ 0.058$ &$-0.861$ &$0.1077$ &$0.0026$ &3 &$ 8.45$ &$0.13$ &$0.16$ &$ 99.6$ &$0.0088$ &Thick \\
LHS1885    &$06^h57^m57^s.0$ &$+62^\circ19\arcmin20\arcsec$ &$1999.0$ &$ 0.327$ &$-0.516$ &$0.0874$ &$0.0023$ &1 &$ 7.71$ &$0.23$ &$0.25$ &$ 52.4$ &$0.0239$ &Mid \\
GJ285      &$07^h44^m40^s.2$ &$+03^\circ33\arcmin09\arcsec$ &$1999.9$ &$-0.348$ &$-0.447$ &$0.1686$ &$0.0027$ &2 &$ 5.72$ &$0.31$ &$0.32$ &$ 2.78$ &$0.0150$ &Thin \\
GJ1101     &$07^h55^m54^s.0$ &$+83^\circ23\arcmin05\arcsec$ &$1999.9$ &$-0.291$ &$-0.604$ &$0.0803$ &$0.0030$ &1 &$ 7.93$ &$0.23$ &$0.25$ &$ 1.11$ &$0.0076$ &Thick \\
G41-10     &$08^h58^m12^s.7$ &$+19^\circ43\arcmin49\arcsec$ &$1998.8$ &$-0.427$ &$ 0.096$ &$0.0332$ &$0.0046$ &1 &$ 9.76$ &$0.25$ &$0.27$ &$  101$ &$0.0082$ &Thick \\
LHS2088    &$08^h59^m56^s.1$ &$+72^\circ57\arcmin36\arcsec$ &$1999.3$ &$ 0.973$ &$-0.035$ &$0.0726$ &$0.0034$ &1 &$ 8.99$ &$0.15$ &$0.18$ &$  138$ &$0.0033$ &Thick \\
LHS267     &$09^h20^m57^s.9$ &$+03^\circ22\arcmin06\arcsec$ &$2000.1$ &$ 0.316$ &$-1.139$ &$0.0608$ &$0.0041$ &1 &$ 8.54$ &$0.23$ &$0.25$ &$  118$ &$0.0027$ &Thick \\
LHS2175    &$09^h42^m23^s.3$ &$+55^\circ59\arcmin02\arcsec$ &$1999.0$ &$-0.703$ &$-0.517$ &$0.0651$ &$0.0087$ &1 &$ 7.55$ &$0.31$ &$0.33$ &$ 72.8$ &$0.0069$ &Thick \\
GJ403      &$10^h52^m04^s.4$ &$+13^\circ59\arcmin51\arcsec$ &$1998.0$ &$-1.115$ &$ 0.198$ &$0.0839$ &$0.0257$ &1 &$ 7.81$ &$0.25$ &$0.27$ &$ 96.8$ &$0.0038$ &Thick \\
G119-62    &$11^h11^m51^s.8$ &$+33^\circ32\arcmin11\arcsec$ &$1998.4$ &$-0.164$ &$ 0.110$ &$0.0683$ &$0.0106$ &1 &$ 7.51$ &$0.31$ &$0.33$ &$ 7.79$ &$0.0197$ &Mid \\
LHS316     &$11^h50^m57^s.9$ &$+48^\circ22\arcmin40\arcsec$ &$1999.0$ &$-1.540$ &$-0.960$ &$0.1221$ &$0.0029$ &3 &$ 7.66$ &$0.17$ &$0.19$ &$  132$ &$0.0042$ &Thick \\
LHS6220    &$12^h05^m29^s.7$ &$+69^\circ32\arcmin23\arcsec$ &$1999.2$ &$-0.447$ &$-0.040$ &$0.0602$ &$0.0134$ &1 &$ 7.91$ &$0.24$ &$0.26$ &$ 96.2$ &$0.0064$ &Thick \\
GJ1156     &$12^h18^m59^s.4$ &$+11^\circ07\arcmin34\arcsec$ &$2000.2$ &$-1.263$ &$ 0.199$ &$0.1529$ &$0.0030$ &3 &$ 7.59$ &$0.14$ &$0.16$ &$0.491$ &$0.0142$ &Thick \\
GJ493.1    &$13^h00^m33^s.5$ &$+05^\circ41\arcmin08\arcsec$ &$2000.2$ &$-0.934$ &$ 0.215$ &$0.1231$ &$0.0035$ &3 &$ 7.68$ &$0.17$ &$0.19$ &$0.600$ &$0.0028$ &Thick \\
LHS2995    &$14^h53^m37^s.2$ &$+11^\circ34\arcmin12\arcsec$ &$2000.3$ &$ 0.076$ &$-0.738$ &$0.0535$ &$0.0041$ &1 &$ 9.67$ &$0.15$ &$0.17$ &$ 88.3$ &$0.0074$ &Thick \\
GJ1204     &$16^h36^m05^s.6$ &$+08^\circ48\arcmin49\arcsec$ &$2000.4$ &$-0.519$ &$-0.151$ &$0.0652$ &$0.0042$ &1 &$ 8.53$ &$0.21$ &$0.23$ &$ 6.41$ &$0.0033$ &Thick \\
NLTT43307  &$16^h40^m06^s.0$ &$+00^\circ42\arcmin19\arcsec$ &$2000.3$ &$ 0.174$ &$-0.163$ &$0.0890$ &$0.0023$ &1 &$ 8.23$ &$0.18$ &$0.20$ &$0.311$ &$0.0043$ &Thin \\
LHS6311    &$16^h40^m20^s.7$ &$+67^\circ36\arcmin05\arcsec$ &$1999.4$ &$-0.262$ &$ 0.361$ &$0.0747$ &$0.0043$ &1 &$ 8.97$ &$0.15$ &$0.17$ &$0.378$ &$0.0091$ &Thick \\
G141-29    &$18^h42^m45^s.0$ &$+13^\circ54\arcmin17\arcsec$ &$1999.4$ &$-0.025$ &$ 0.347$ &$0.0933$ &$0.0115$ &1 &$ 7.57$ &$0.15$ &$0.17$ &$ 8.07$ &$0.0132$ &Thick \\
LHS3461    &$19^h24^m16^s.3$ &$+75^\circ33\arcmin12\arcsec$ &$2000.4$ &$ 0.374$ &$ 0.591$ &$0.0903$ &$0.0051$ &1 &$ 8.98$ &$0.12$ &$0.15$ &$  114$ &$0.0121$ &Thick \\
GJ1243     &$19^h51^m09^s.3$ &$+46^\circ29\arcmin00\arcsec$ &$1998.4$ &$ 0.188$ &$ 0.266$ &$0.0841$ &$0.0024$ &1 &$ 7.79$ &$0.24$ &$0.26$ &$0.593$ &$0.0085$ &Mid \\
GJ791.2AB\tablenotemark{j} &$20^h29^m48^s.4$ &$+09^\circ41\arcmin20\arcsec$ &$2000.4$ &$ 0.665$ &$ 0.132$ &$0.1138$ &$0.0019$ &3 &$ 7.33$ &$0.21$ &$0.23$ &$0.346$ &$0.0038$ &Mid \\
GJ856A     &$22^h23^m29^s.0$ &$+32^\circ27\arcmin33\arcsec$ &$1998.5$ &$ 0.255$ &$-0.208$ &$0.0622$ &$0.0100$ &2 &$ 6.07$ &$0.25$ &$0.27$ &$0.854$ &$0.0070$ &Thin \\
LHS543a    &$23^h25^m40^s.2$ &$+53^\circ08\arcmin06\arcsec$ &$1998.9$ &$ 0.986$ &$ 0.328$ &$0.0404$ &$0.0031$ &1 &$ 9.00$ &$0.28$ &$0.30$ &$ 24.8$ &$0.0077$ &Thick \\
LHS3978    &$23^h35^m41^s.3$ &$+06^\circ11\arcmin21\arcsec$ &$2000.7$ &$ 0.566$ &$ 0.268$ &$0.0416$ &$0.0032$ &1 &$10.19$ &$0.15$ &$0.18$ &$ 1.67$ &$0.0055$ &Thick \\
\enddata

\tablenotetext{a}{Coordinates from 2MASS, epoch of measurement as specified in the ``epoch'' column.}
\tablenotetext{b}{From \citet{ls2005}.}
\tablenotetext{c}{Reference for parallax measurement: (1) Yale Parallax Catalog, \citet{vanaltena1995}; (2) The {\it Hipparcos} catalog, \citet{perryman1997}; (3) NASA NStars database ({\tt http://nstars.arc.nasa.gov/}); (4) \citet{dahn2002}.  Compiled by \citet{lepine2005}. }
\tablenotetext{d}{From 2MASS, converted to the CIT system of \citet{elias1982,elias1983} using the transformation from the 2MASS explanatory supplement.}
\tablenotetext{e}{Using the $K$-band absolute magnitude and the relation of \citet{delfosse2000}.}
\tablenotetext{f}{Derived from $M$ using a polynomial fit to the empirical mass-radius data of \citet{ribas2006}.}
\tablenotetext{g}{While we have not attempted to estimate formal
  period uncertainties, our simulations indicate that the uncertainty
  in the periods is $< 1\%$ for $P < 10\ {\rm d}$, $\approx 1\%$ for $10
  \le P < 20\ {\rm days}$, $2\%$ for $20 \le P < 50\ {\rm days}$,
  $5-10\%$ for $50 \le P < 100\ {\rm days}$, and $20-30\%$ for $P \ge
  100\ {\rm days}$.}
\tablenotetext{h}{Semi-amplitude of modulation seen in the MEarth bandpass.}
\tablenotetext{i}{Kinematic population, see \S \ref{kin_sect}.}
\tablenotetext{j}{Visual binary, unresolved in these observations; see \citet{benedict2000}.}

\end{deluxetable*}
\clearpage
\end{landscape}

\end{document}